\documentclass{aastex631}
\let\tablenum\relax
\usepackage{siunitx}
\usepackage{amsfonts,mathbbol}
\usepackage{physics}
\usepackage{pifont}
\usepackage{amsmath}

\newcommand{\chandra}{\textit{Chandra}\ }
\DeclareSIUnit\erg{erg}
\DeclareSIUnit\parsec{pc}
\DeclareSIUnit\jansky{Jy}
\DeclareSIUnit\beam{beam}
\DeclareSIUnit\year{yr}
\DeclareSIUnit\mas{mas}

\shorttitle{A central radio source in 47 Tuc}
\shortauthors{Paduano et al.}
\graphicspath{{./}{figures/}}
\begin{document}

\title{Ultra-deep ATCA imaging of 47 Tucanae reveals a central compact radio source}

\correspondingauthor{Arash Bahramian}
\email{arash.bahramian@.curtin.edu.au}

\author[0000-0003-0725-8330]{Alessandro Paduano}
\affiliation{International Centre for Radio Astronomy Research - Curtin University \\
1 Turner Ave, Bentley, WA 6102, Australia}

\author[0000-0003-2506-6041]{Arash Bahramian}
\affiliation{International Centre for Radio Astronomy Research - Curtin University \\
1 Turner Ave, Bentley, WA 6102, Australia}

\author[0000-0003-3124-2814]{James C. A. Miller-Jones}
\affiliation{International Centre for Radio Astronomy Research - Curtin University \\
1 Turner Ave, Bentley, WA 6102, Australia}

\author[0000-0002-4485-6471]{Adela Kawka}
\affiliation{International Centre for Radio Astronomy Research - Curtin University \\
1 Turner Ave, Bentley, WA 6102, Australia}

\author[0000-0002-2801-766X]{Tim J. Galvin}
\affiliation{International Centre for Radio Astronomy Research - Curtin University \\
1 Turner Ave, Bentley, WA 6102, Australia}
\affiliation{CSIRO Space \& Astronomy\\
PO Box 1130, Bentley, WA 6102, Australia}

\author[0000-0002-9396-7215]{Liliana Rivera Sandoval}
\affiliation{Department of Physics and Astronomy, University of Texas Rio Grande Valley \\
Brownsville, TX 78520, USA}

\author[0000-0001-6604-0505]{Sebastian Kamann}
\affiliation{Astrophysics Research Institute, Liverpool John Moores University \\ 
146 Brownlow Hill, Liverpool L3 5RF, UK}

\author[0000-0002-1468-9668]{Jay Strader}
\affiliation{Center for Data Intensive and Time Domain Astronomy, Department of Physics and Astronomy, Michigan State University \\
East Lansing, MI 48824, USA}

\author[0000-0002-8400-3705]{Laura Chomiuk}
\affiliation{Center for Data Intensive and Time Domain Astronomy, Department of Physics and Astronomy, Michigan State University \\
East Lansing, MI 48824, USA}

\author[0000-0003-3944-6109]{Craig O. Heinke}
\affiliation{Department of Physics, University of Alberta \\
CCIS 4-181, Edmonton, AB T6G 2EI, Canada}

\author[0000-0003-0976-4755]{Thomas J. Maccarone}
\affiliation{Department of Physics \& Astronomy, Texas Tech University \\
Box 41051, Lubbock, TX 79409-1051, USA}

\author[0000-0001-6187-5941]{Stefan Dreizler}
\affiliation{Institut f\"ur Astrophysik, Georg-August-Universit\"at G\"ottingen\\ Friedrich-Hund-Platz 1, 37077 G\"ottingen, Germany}



\begin{abstract}
We present the results of an ultra-deep radio continuum survey, containing $\sim480$ hours of observations, of the Galactic globular cluster 47 Tucanae with the Australia Telescope Compact Array. This comprehensive coverage of the cluster allows us to reach RMS noise levels of \SI{1.19}{\micro\jansky\per\beam} at 5.5 GHz, \SI{940}{\nano\jansky\per\beam} at 9 GHz, and \SI{790}{\nano\jansky\per\beam} in a stacked 7.25 GHz image. This is the deepest radio image of a globular cluster, and the deepest image ever made with the Australia Telescope Compact Array. We identify ATCA J002405.702-720452.361, a faint ($6.3\pm1.2$ \si{\micro\jansky} at 5.5 GHz, $5.4\pm0.9$ \si{\micro\jansky} at 9 GHz), flat-spectrum ($\alpha=-0.31\pm0.54$) radio source that is positionally coincident with the cluster centre and potentially associated with a faint X-ray source. No convincing optical counterpart was identified. We use radio, X-ray, optical, and UV data to show that explanations involving a background active galactic nucleus, a chromospherically active binary, or a binary involving a white dwarf are unlikely. The most plausible explanations are that the source is an undiscovered millisecond pulsar or a weakly accreting black hole. If the X-ray source is associated with the radio source, the fundamental plane of black hole activity suggests a black hole mass of $\sim54-6000$ M$_{\odot}$, indicating an intermediate-mass black hole or a heavy stellar-mass black hole.

%Tonight, Alessandro spends 400 hours staring at a single source, Arash makes some plots, and Craig thinks it's the brightest active binary, in the world.
\end{abstract}


%%%%%%%%%%%%%%%%%%%%%%%%%%%%%%%%%%%%%%%%%%%%%%%%%%%%%%%%%%%%%%%%%%%%%%%%%%%%%%%%%%%%%%%%%%
\section{Introduction} \label{sec:intro}
\subsection{Multiwavelength sources in globular clusters}
The Milky Way contains 158 known globular clusters \citep[GCs, ][]{Baumgardt2019}; large, gravitationally bound clusters of stars that orbit the Galactic Centre. When compared to the Galactic field, it has been shown that X-ray binaries (XRBs), binary systems containing a black hole (BH) or a neutron star (NS), are overabundant in GCs when compared to the Galactic field \citep{Clark1975}. This overabundance is due to the additional dynamical formation channels of XRBs in GCs \citep[e.g., ][]{Sutantyo1975,Fabian1975,Hills1976}. 

It has been known for several decades that GCs contain a high number of X-ray emitting sources \citep{Clark1975a}. These initial detections spurred the first radio surveys of GCs using the NRAO interferometer \citep{Johnson1976,Johnson1977,Rood1978}, the Arecibo 305-m telescope \citep{Terzian1977}, and the 100-m Effelsberg radio telescope. These initial radio surveys spanned frequency ranges of $\sim2-11$ GHz, and were sensitive to bright sources with flux densities $>\SI{1}{\milli\jansky}$. The discovery of millisecond pulsars \citep[MSPs, ][]{Alpar1982} also spawned further radio timing surveys of GCs at lower frequencies to search for pulsars \citep{Hamilton1985,Fruchter1990,Fruchter2000}. Pulsars are abundant in GCs. To date, 257 pulsars are known in 36 GCs\footnote{\url{https://www3.mpifr-bonn.mpg.de/staff/pfreire/GCpsr.html}}, with Terzan 5 and 47 Tucanae containing the largest number of pulsars. Bayesian estimates indicate that there are potentially more than 2000 pulsars within Galactic GCs \citep{Turk2013}.

Pulsars are not the sole class of radio sources that are expected to be detected in radio continuum imaging. Accreting XRBs, abundant in GCs, are expected to be visible at radio frequencies due to the radio synchrotron emission associated with non-thermal jets. Additionally, GCs are potentially hosts for the formation of intermediate-mass black holes (IMBHs) through a variety of different formation channels such as through sequential mergers of stellar-mass BHs \citep{Miller2002a}, or through the runaway growth of a massive object through stellar collisions \citep{PortegiesZwart2002,PortegiesZwart2004}. See \citet{Greene2020} for a recent review on IMBHs. This makes GCs prime targets in the search for IMBHs. An IMBH in a GC is assumed to accrete a portion of the gas within its sphere of influence, producing radio or X-ray emission that is potentially detectable \citep{Maccarone2004}. 

Due to the correlation among radio luminosity, X-ray luminosity, and mass (the fundamental plane of BH activity, \citealt{Merloni2003,Falcke2004,Miller-Jones2012,Plotkin2012}), the radio emission from an IMBH is expected to be brighter (at a given X-ray luminosity) than that from a stellar-mass BH. This makes radio continuum searches one of the best ways to try and detect IMBHs in GCs. Interpreting these searches involves making assumptions as to the radiative efficiency of any accretion onto an IMBH and the expected accretion rates \citep{Pellegrini2005}, in addition to the expected gas density in GCs based on measurements using pulsar dispersion measures in 47 Tucanae \citep{Freire2001,Abbate2018}. Several papers have searched for IMBH accretion signatures in Galactic GCs \citep[e.g., ][]{Maccarone2005,Maccarone2008,Cseh2010,Lu2011,Strader2012,Tremou2018,Su2022}, with the most recent limits indicating that either IMBHs with masses $\gtrsim1000$~M$_{\odot}$ are not present in GCs, the accretion onto central IMBHs is more inefficient than predicted by \citet{Maccarone2003}, or that the gas density is lower in most clusters than in 47 Tucanae.

Recently, the first radio continuum imaging survey reaching mean noise levels $<10$ \si{\micro\jansky\per\beam} of 50 GCs in the Milky Way was conducted. The MAVERIC (Milky Way ATCA VLA Exploration of Radio Sources in Clusters) survey \citep{Shishkovsky2020,Tudor2022} was a systematic survey of 50 Galactic GCs with the Karl G. Jansky Very Large Array (VLA) and the Australia Telescope Compact Array (ATCA) to assess the associated radio source populations. The pilot surveys revealed BH candidates in M22 \citep{Strader2012a}, M62 \citep{Chomiuk2013}, and 47 Tucanae \citep{Miller-Jones2015}, with the full survey detecting additional BH candidates in M10 \citep{Shishkovsky2018} and NGC~6397 \citep{Zhao2020}, and transitional millisecond pulsar candidates in Terzan 5 \citep{Bahramian2018a} and NGC~6652 \citep{Paduano2021}. A new ``hidden'' MSP was also detected in NGC~6397 \citep{Zhao2020}, which has recently been confirmed by MeerKAT and Parkes timing \citep{Zhang2022}. As alluded to above, the full MAVERIC survey has been used to search for signatures of IMBH accretion in GCs \citep{Strader2012,Tremou2018}. No IMBH signatures were detected in any GC in the MAVERIC sample.

\subsection{47 Tucanae}\label{sec:intro47tuc}
47 Tucanae (47 Tuc, NGC 104) is the second-brightest globular cluster after $\omega$-Centauri. 47 Tuc has a mass of $\sim\num{8.95e5}$~M$_{\odot}$, and is located at a distance of $4.52\pm0.03$ kpc \citep{Baumgardt2021}, with a very low extinction of $E(B-V)=0.04\pm0.02$ \citep{Salaris2007}. These last two points make the cluster an easy target for multi-wavelength studies, with 47 Tuc being studied extensively by the \textit{Chandra} X-ray Observatory and Hubble Space Telescope (HST). 47 Tuc contains a rich population of over 300 known X-ray sources of various classes, including low-mass X-ray binaries, cataclysmic variables (CVs), MSPs, and chromospherically active binaries (ABs) \citep{Heinke2005,Bhattacharya2017}. Of particular interest for this work is the X-ray source CXOGlb J002405.6$-$720452, also known as [GHE2001] W286 (hereafter W286, $\alpha=00:24:05.697$, $\delta=-72:04:52.306$, \citealt{Heinke2005,Bhattacharya2017,Su2022}), which falls within \SI{1}{\arcsecond} of the photometric centre of 47 Tuc (as measured by \citealt{Goldsbury2010}). A potential optical counterpart was suggested for this source by \citet{Edmonds2003}, corresponding to a BY Draconis variable Cl* NGC 104 EGG V32 (hereafter PC1-V32, $\alpha=00:24:05.404$, $\delta=-72:04:52.316$) first identified by \citet{Albrow2001}, with a period of 1.64 days. BY Draconis (BY Dra) sources are main sequence variable stars which exhibit luminosity variations due to chromospheric activity and the rotation of the star.

47 Tuc contains 29 known pulsars at the time of writing, which is the second largest number of pulsars in a GC behind Terzan 5. Nineteen pulsars are in binary systems, and 23 pulsars have phase-coherent timing solutions. The majority of the pulsars in 47 Tuc were discovered prior to 2003, with nine further sources identified and confirmed in the years since. Modelling by \citet{Ye2021} indicates that 47 Tuc may contain $\sim$50 MSPs, meaning there are potentially still several MSPs yet to be discovered in the cluster.

The existence of an IMBH in 47 Tuc has never been proven in previous literature. \citet{Freire2017} and \citet{Abbate2018} indicate that based on the pulsar accelerations in the cluster, a central IMBH is not needed, with an upper-limit on the mass of a central IMBH of $\sim4000$~M$_{\odot}$ \citep{Abbate2018}. Velocity dispersion modelling by \citet{Mann2019} found that the velocity dispersion in the core of the cluster can be produced by the binaries and BHs in the core, such that an IMBH is not needed to explain the velocity dispersion. They found an IMBH mass of $40\pm1650$~M$_{\odot}$, and that a central IMBH is only needed if the retention fraction of stellar-mass BHs and NSs is very low. Further multimass modelling by \citet{Henault-Brunet2020} also indicates that an IMBH is not required in 47 Tuc to explain various observational constraints. Additional modelling of the cluster by \citet{Ye2021} indicated approximately 200 BHs could be present, giving a total mass of BHs in 47 Tuc of $\sim2000$~M$_{\odot}$. Most recently, \citet{DellaCroce2023} analyzed the kinematics of the cluster's central region and inferred that the observed kinematics are inconsistent with a central IMBH more massive than $578$~M$_{\odot}$.

The sheer number of X-ray sources in 47 Tuc makes it a very appealing target for radio continuum surveys. The first of these surveys was conducted by \citet{McConnell2000} with the ATCA, reaching RMS noise levels of 42 and \SI{46}{\micro\jansky\per\beam} at 1.4 and 1.7 GHz respectively, enabling the detection of 11 radio sources within 5' of the cluster centre. These 11 sources included the detections of two pulsars with known positions. This initial survey was built upon by \citet{Fruchter2000}, who presented images with RMS noise levels of \SI{32}{\micro\jansky\per\beam}. The initial survey by \citet{McConnell2000} was extended to include 170 hr of ATCA data the following year, pushing the RMS noise down to \SI{18}{\micro\jansky\per\beam} \citep{McConnell2001} and detecting nine radio sources in the cluster core. Following the bandwidth upgrade to the ATCA \citep{Wilson2011}, \citet{Lu2011} obtained approximately 18 hr of ATCA data at 5.5 and 9 GHz in 2010, reaching an RMS noise level of \SI{13.3}{\micro\jansky\per\beam} after stacking both bands. These observations were subsequently combined with the MAVERIC observations of 47 Tuc, reaching RMS noise levels of 4.4 and \SI{5.7}{\micro\jansky\per\beam} at 5.5 and 9 GHz respectively, and used to identify the BH candidate X9 \citep{Miller-Jones2015}. These radio continuum surveys have also contributed to placing mass upper limits on central IMBHs, with \citet{Lu2011} and \citet{Tremou2018} using the fundamental plane of black hole activity to place $3\sigma$ upper mass limits of $520-4900$~M$_{\odot}$ and $1040$~M$_{\odot}$ respectively. While both studies discussed the significant uncertainties associated with mass estimates derived from the fundamental Plane, they did not account for this scatter \citep[recently quantified as 1\,dex by][]{Gultekin2019} in their quoted mass limits.

In this paper, we combine the archival observations of \citet{Lu2011} and \citet{Miller-Jones2015} with over 400 hr of new ATCA data to make the deepest radio image of a GC to date. Our ultra-deep campaign reaches next-generation RMS noise levels of $\sim\SI{790}{\nano\jansky\per\beam}$, representing the deepest radio image ever made with the ATCA. The unparalleled depths that this imaging campaign reaches have allowed for the detection of a faint radio source (ATCA J002405.702-720452.361) at the photometric centre of 47 Tuc. This paper will present an investigation of this radio source, and a follow-up paper will present the full radio source catalogue from this campaign. In Section~\ref{sec:method}, we describe our radio observations and data reduction, in addition to other data analysed during this study. The results are presented in Section~\ref{sec:results}. In Section~\ref{sec:discussion}, we provide a discussion of our findings and step through the possible source classes for ATCA J002405.702-720452.361. In Section~\ref{sec:conclusion} we present our conclusions.

%%%%%%%%%%%%%%%%%%%%%%%%%%%%%%%%%%%%%%%%%%%%%%%%%%%%%%%%%%%%%%%%%%%%%%%%%%%%%%%%%%%%%%%%%%
\section{Observations and data reduction} \label{sec:method}
\subsection{ATCA observations}
47 Tuc was observed by the ATCA under the project code C3427 over 41 epochs between 2021 March 31 and 2022 May 6. For all but five epochs, the array was in an extended 6-km configuration. This array configuration was chosen to maximise spatial resolution. For observations on 2021 December 28, 2021 December 30, 2022 January 2, and 2022 April 25, the array was in the 1.5-km configuration, and for 2022 May 6 the array was in the 750-m configuration. This was done to obtain shorter baseline coverage to improve our sensitivity to some extended sources in the field. A full overview of the date, duration, and array configuration of each epoch is shown in Table~\ref{tab:observations}. 

Observations were conducted in two bands simultaneously, using the Compact Array Broadband Backend (CABB) correlator \citep{Wilson2011}. The two bands each had a bandwidth of 2048 MHz, split into equal 1 MHz channels, and were centred on frequencies of 5.5 and 9 GHz. The source B1934-638 was used as the primary calibrator for bandpass and flux calibration, and the source B2353-686 was used as the secondary calibrator for amplitude and phase calibration. Occasionally, the source J0047$-$7530 was used as the secondary calibrator for times where B2353-686 had set. During each observation, after initial calibration on B1934-638 for approximately 15 minutes, we cycled between observing the secondary calibrator and target for 1 minute and 15 minutes respectively. During poorer observing conditions, the target integration time was reduced to 5 minutes between secondary calibrator scans.

    \begin{table*}
        \centering
        \caption{The ATCA observations of 47 Tuc taken under the project code C3427. For each epoch, the date and start time (in UTC), integration time, and array configuration are given.}
        \label{tab:observations}
        \begin{tabular}{lccc}
            \hline
            \hline
            Date & Start time (UTC) & Integration time (hr) & Array configuration \\
            \hline
            2021-03-31 & 21:00 & 11.66 & 6D \\
            2021-04-01 & 20:00 & 11.56 & 6D \\
            2021-04-05 & 22:00 & 9.63 & 6D \\
            2021-04-07 & 01:30 & 9.12 & 6D \\
            2021-04-08 & 21:30 & 10.87 & 6D \\
            2021-04-09 & 20:30 & 11.54 & 6D \\
            2021-04-11 & 20:00 & 11.58 & 6D \\
            2021-04-14 & 17:00 & 11.53 & 6D \\
            2021-04-15 & 20:00 & 11.45 & 6D \\
            2021-04-16 & 17:00 & 11.50 & 6D \\
            2021-04-17 & 20:00 & 11.69 & 6D \\
            2021-04-23 & 23:00 & 10.00 & 6D \\
            2021-06-29 & 18:30 & 9.72 & 6B \\
            2021-06-30 & 17:30 & 10.83 & 6B \\
            2021-07-01 & 17:30 & 10.62 & 6B \\
            2021-09-09 & 06:00 & 11.52 & 6A \\
            2021-09-12 & 09:00 & 11.49 & 6A \\
            2021-09-17 & 05:00 & 11.20 & 6A \\
            2021-09-19 & 12:00 & 11.24 & 6A \\
            2021-09-20 & 10:00 & 11.35 & 6A \\
            2021-09-21 & 08:30 & 11.48 & 6A \\
            2021-09-22 & 08:30 & 11.56 & 6A \\
            2021-09-23 & 08:30 & 12.40 & 6A \\
            2021-09-25 & 07:30 & 11.62 & 6A \\
            2021-09-26 & 09:00 & 11.36 & 6A \\
            2021-09-28 & 06:00 & 11.97 & 6A \\
            2021-09-29 & 08:30 & 9.64 & 6A \\
            2021-10-01 & 07:30 & 11.68 & 6A \\
            2021-11-17 & 13:00 & 2.75 & 6C \\
            2021-11-22 & 07:00 & 8.91 & 6C \\
            2021-12-28 & 06:00 & 7.68 & 1.5A \\
            2021-12-30 & 02:00 & 11.75 & 1.5A \\
            2022-01-02 & 02:00 & 10.11 & 1.5A \\
            2022-01-22 & 00:30 & 11.68 & 6A \\
            2022-01-26 & 00:00 & 5.31 & 6A \\
            2022-01-27 & 00:00 & 11.41 & 6A \\
            2022-01-28 & 00:00 & 11.96 & 6A \\
            2022-01-30 & 00:00 & 12.82 & 6A \\
            2022-01-30 & 23:30 & 13.81 & 6A \\
            2022-04-25 & 18:00 & 11.53 & 1.5A \\
            2022-05-06 & 15:30 & 13.64 & 750D \\
            \hline
        \end{tabular}
    \end{table*}

\subsection{Radio data reduction and imaging} \label{sec:reduction}
We reduced the data for each band separately. Data calibration was performed using standard procedures in \textsc{miriad} \citep{Sault1995}, before we imported the \textit{uv}-visibilities into the Common Astronomy Software Application \citep[\textsc{casa};][]{McMullin2007} for imaging. We used the \verb#tclean# task for imaging, and used a robust weighting factor of 1.0 to provide a good balance between image sensitivity and resolution. We used the multi-term multi-frequency synthesis deconvolver with two Taylor terms to account for the frequency-dependence of sources in the field, and the resulting images were primary beam corrected using the task \verb#impbcor#. We used cell sizes of \SI{0.3}{\arcsecond} and \SI{0.15}{\arcsecond} and image sizes of 3072 and 5625 pixels for imaging the 5.5 and 9 GHz bands respectively. The image centre was set to approximately \SI{1}{\arcminute} to the south of the cluster centre. These image sizes and this image phase-centre offset were applied to aid in the reconstruction and deconvolution of some bright, extended sources towards the edge of the image field at 5.5 GHz.

Primary beam corrected images were made at both 5.5 and 9 GHz. We also imaged a co-stack of these two bands, which resulted in an image with an apparent central frequency of 7.25 GHz and a lower image noise than the separate bands. To create the deepest possible images we stacked all 5.5 and 9 GHz epochs, excluding the data taken on 2021 April 7 and 2021 September 29 due to poor observing conditions. Again, we stacked and imaged both bands to produce a deep 7.25 GHz image of the field.

To verify that our imaging techniques in \textsc{casa} were producing reliable results, we also imaged our data using \textsc{ddfacet} \citep{Tasse2018}. \textsc{ddfacet} performs spectral deconvolution using image-plane faceting. Given the small field-of-view of this campaign and to improve computational efficiency, we imaged with four facets in a $2\times2$ configuration. All other imaging parameters such as robustness, image size and cell size were identical to those used in \textsc{casa}. The images produced using \textsc{ddfacet} contained similar source distributions and noise structure, giving us confidence that our images were the correct representation of the data\footnote{Furthermore, we found that the flux and spectral slope of the main target of this work (ATCA J002405.702-720452.361; see \S\ref{sec:results}) were consistent between the two methods.}. 

\subsubsection{Archival radio data}
To complement our survey, we searched the ATCA archive for previous observations of 47 Tuc. We combined our data with data taken by \citet{Lu2011}, \citet{Miller-Jones2015}, and \citet{Bahramian2017}, giving us an extra $\sim35$ hr of on-target observations. These data were reduced in the same manner described in Section~\ref{sec:reduction}. The inclusion of these data allowed us to reach RMS noise levels of \SI{1.19}{\micro\jansky\per\beam} at 5.5 GHz and \SI{940}{\nano\jansky\per\beam} at 9 GHz. When we stacked all the available data at 5.5 and 9 GHz together to
make an image with an effective frequency of 7.25 GHz, the RMS noise was \SI{790}{\nano\jansky\per\beam}, representing the deepest image made to date with the ATCA.

\subsection{X-ray data}
We used the X-ray source catalogue of 47 Tuc compiled by \citet{Bhattacharya2017} as our main source catalogue to search for potential X-ray counterparts to radio sources. To investigate the X-ray properties of 47 Tuc beyond the scope of \citet{Bhattacharya2017}, we queried the \chandra archive for previous X-ray observations of the cluster using the \chandra/ACIS detector. Nineteen observations of 47 Tuc were made using the \chandra/ACIS detector between 2000 and 2015, totalling more than 500 ks of data which we obtained for our analysis.

For data reprocessing and analysis, we used \textsc{ciao} 4.14 with \textsc{caldb} 4.9.7 \citep{Fruscione2006}. All X-ray data were reprocessed using \verb#chandra_repro# before stacking. To stack the observations, we first corrected the coordinate system of each observation by using \verb#wavdetect# for source detection, and then \verb#wcs_match# and \verb#wcs_update# to create a matrix transformation to apply and correct the WCS parameters of each observation based on a single reference observation. When correcting the WCS parameters for stacking, we were only interested in the relative astrometry between each observation as our main aim for this X-ray analysis was to extract X-ray spectra. Considerations of the absolute astrometry are outlined in Section~\ref{sec:method_astrometry}. We used the task \verb#merge_obs# to stack the observations. 

Given that 47 Tuc had not been observed by \chandra since 2015 prior to our ATCA campaign, we obtained new \chandra data of the cluster to search for signs of significant X-ray variability. Under Director's Discretionary Time (DDT), 47 Tuc was observed for \SI{9.62}{ks} on 2022 January 26 (Obs ID: 26229) and for \SI{9.83}{ks} on 2022 January 27 (Obs ID: 26286), giving us almost \SI{20}{ks} of new \chandra data of 47 Tuc for the first time since 2015. These data were reprocessed via the same method described above, and also stacked with the archival \chandra data.

\subsection{HST Data}
We used optical data from a variety of different sources in order to complement our survey and search for potential optical counterparts to radio sources detected. Primarily, we used data of 47 Tuc from the HST UV Globular Cluster Survey \citep[HUGS, ][]{Piotto2015,Nardiello2018}, to get an initial insight into the positions and properties of optical sources in the cluster. For further analysis, we used optical and UV data taken with the HST, specifically using the Space Telescope Imaging Spectrograph (STIS). STIS data were obtained from the Mikulski Archive for Space Telescopes (MAST). We used data under the program ID 8219 (PI: Knigge; \citealt{Knigge2002}) which were obtained between 1999 September 10 and 2000 August 16. Primarily, we looked at far-ultraviolet (FUV) data, which used the MIRFUV filter with the FUV-MAMA detector.

In this work, we also used near-ultraviolet (NUV) and optical data from the General Observer programs 12950 and 9281. The NUV dataset was taken on 2013 August 13 and contains images in the F390W and F300X filters. The optical images were acquired over three visits on 2002 September 20, 2002 October 2/3 and 2002 October 11 in the F435W (B), F625W (R) and F658N (H$\alpha$) filters. All images were calibrated and astrometrically corrected to the epoch J2000 as described in \citet{RiveraSandoval2015}, with PSF photometry carried out as mentioned in \citet{RiveraSandoval2015,RiveraSandoval2018}. The optical data were aligned and photometrically analysed using using the software \textsc{DOLPHOT} \citep{Dolphin2016}. Photometry was obtained using the individual FLC images in the three filters simultaneously, and we used the combined DRC image in the B filter as a reference frame. The magnitude limits are magnitude 27 for the NUV data, magnitude 24.5 for the R and H$\alpha$ data, and magnitude 25 for the B data.

\subsection{MUSE integral field spectroscopy}
47~Tuc was observed with MUSE \citep{Bacon2010} in narrow-field mode (NFM) during the night of 2019-11-01, for a total exposure time of 4$\times$600~s. In NFM, MUSE provides a field of view of 7.5$\times$7.5~arcsec with a spatial sampling of 0.025~arcsec and uses the GALACSI module \citep{Strobele2012} in laser tomographic adaptive optics (LTAO) mode to achieve a spatial resolution of $\lesssim$0.1~arcsec. The spectral coverage is from 470 to 930~nm with a constant full-width at half maximum (FWHM) of 2.5~$\text{\AA}$, corresponding to a spectral resolution of R$\sim$1700-3500.  The observations were taken as part of the MUSE GTO survey of globular clusters (PI: Kamann/Dreizler), described in \citet{Kamann2018}.

The data were reduced with version 2.8.1 of the standard MUSE pipeline \citep{Weilbacher2020}. The pipeline performs the basic reduction steps (such as bias subtraction, flat fielding, wavelength calibration, or flux calibration) on each individual exposure in order to create a pixtable, which contains the WCS coordinates, wavelength, flux, and flux uncertainty of every valid CCD pixel. In the last step, the individual pixtables are combined and resampled to the final data cube. 

We extracted individual spectra from the data cube using \textsc{PampelMuse} \citep{Kamann2013}. \textsc{PampelMuse} uses a reference catalog of sources in the observed field in order to measure the positions of the resolved stars and the point spread function (PSF) as a function of wavelength. This information is then used in order to deblend the spectra of the individual stars from the cube. The reference catalog used in the analysis was published by \citet{Anderson2008} and is based on the HST/ACS survey of Galactic globular clusters presented in \citet{Sarajedini2007}. In order to model the non-trivial shape of the MUSE PSF in NFM, we used the \textsc{MAOPPY} model by \citet{Fetick2019}.

The extracted spectra were analysed as outlined in \citet{Husser2016}. In particular, we measured stellar radial velocities from the extracted spectra by first cross-correlating each of them against a synthetic template spectrum with matched stellar parameters and then performing a full-spectrum fit.  The templates used to perform the cross correlation as well as the full spectrum fitting were taken from the \textsc{GLib} library presented in \citet{Husser2013}. During the full-spectrum fitting, we also fitted for the effective temperature $T_{\rm eff}$ and the metallicity [M/H] of each star.  The initial values for these parameters and the surface gravity $\log g$ (which was fixed during the analysis) were obtained from a comparison of the HST photometry available in the reference catalog and isochrones from the \citet{Bressan2012} database. 

In order to search for any resolved H$\alpha$ emission, we further created a residuals map from the MUSE data in the wavelength range around 656.3~nm. To do so, we first subtracted the contribution of each resolved star using its spectrum, position in the cube, and the \textsc{MAOPPY} PSF model valid for each wavelength step. In order to suppress any artefacts from the extraction process (like PSF residuals or faint stars missing in the catalog), we also created the residuals map for two wavelength ranges bluewards and redwards of H$\alpha$. The two off-band residual maps were averaged and subtracted from the on-band residual map.

\subsection{Astrometry} \label{sec:method_astrometry}
As we are combining data that have been taken several years apart, we need to consider the epochs that these data were taken in and shift the epochs in a manner such that the coordinates of different surveys can be compared. All source positions and surveys that we consider have coordinates in the equinox J2000. 

The epoch of the radio data ranges from J2010.07 to J2022.34. To compare the coordinates of radio sources to other surveys from different epochs, we take the epoch of the radio data to be J2021.2, which corresponds to an average of all observation epochs weighted by the respective integration times.

The positions of the X-ray sources from \citet{Bhattacharya2017} were aligned in that paper with those of pulsars in the cluster based on \citet{Freire2003} and \citet{Freire2017}, which are given in the epoch MJD 51600 ($\sim$J2000.16). The positions of optical sources from the HUGS survey are based on the epoch of \textit{Gaia} DR1, which is J2015.0 \citep{Nardiello2018}. The position of the cluster centre has been adopted from \citet{Goldsbury2010} for this work. This position has been astrometrically corrected to the 2MASS \citep{Skrutskie2006}, but does not appear to have been shifted to a particular epoch. Thus, the astrometry is likely to be the epoch at which the data were taken, $\sim$J2006.2.

To check the astrometric frame of the ATCA data, we identified 11 millisecond pulsars (PSR C, D, E, J, L, M, O, Q, S, T, U) whose positions and proper motions have been measured at high precision using pulsar timing observations \citep{Freire2017}, which are detected at $>3\sigma$ in the ATCA dataset, and which are not confused with other sources. We transformed the timing positions from their measured epoch (J2000.16) to the adopted ATCA epoch (J2021.2). The median offsets in
right ascension and declination, in the sense PSR--ATCA, are $-0.08\pm0.05$" and $+0.04\pm0.06$", respectively, where the uncertainties listed are the standard errors of the mean. The weighted mean offsets in right ascension and declination are consistent with these values, but slightly smaller, at $-0.05\pm0.03$" and $+0.01\pm0.05$",
where these are the uncertainties in the weighted means. The rms offsets in each coordinate are fully consistent with the median uncertainties in the ATCA positions of this sample. These comparisons suggest that (i) there is no evidence for an offset between the ATCA astrometry and the precise frame of the pulsars, and (ii) from this comparison alone, any offset
is limited to $0.16\arcsec$ at the $2\sigma$ level.

Separately, we checked the absolute astrometry of the HUGS optical data by cross-checking with \emph{Gaia} DR3, finding no evidence for an offset and an rms scatter of only $\sim 0.01$" for bright stars. We also compared the HUGS positions of the four millisecond pulsars (S, T, U, V) with known optical counterparts \citep{RiveraSandoval2015} that are present in HUGS, finding very consistent astrometry between HUGS and the proper-motion corrected pulsar positions, with no evidence for an offset and an rms $\lesssim 0.01$". Together these checks show that the optical and radio frames are aligned to within a fraction of an HST pixel.

We re-checked the X-ray astrometry using the millisecond pulsars with precise positions and which \citet{Bhattacharya2017} identify as being relatively uncrowded in the X-ray (pulsars C, D, E, H, J, M, N, O, Q, T, U, W, Y, Z, ab). In right ascension, there is indeed no offset (PSR--Chandra = $0.00\pm0.01''$), but we find evidence for a small offset in declination of $-0.08\pm0.02''$. The overall rms scatter in the PSR--Chandra coordinates is 0.07". This is much smaller than the formal 95\% astrometric confidence intervals of the \chandra X-ray coordinates of the pulsars, which range from 0.30--0.35$''$ \citep{Bhattacharya2017}. Hence the uncertainty in the absolute astrometric frame of the X-ray data does not meaningfully affect our results. Nonetheless, given that the declination offset is formally significant, we do apply this offset to the position of the source W286 discussed below.

For the remainder of the paper, to convert source positions to different epochs for comparison, we assume that the sources that are associated with the cluster move with the cluster proper motion of $\mu_{\alpha}=\SI{5.25}{\mas\per\year}$ and $\mu_{\delta}=\SI{-2.53}{\mas\per\year}$. These proper motion values have been adopted from \citet{Baumgardt2019}, who derive the mean proper motions of Galactic GCs from \textit{Gaia} DR2.

%%%%%%%%%%%%%%%%%%%%%%%%%%%%%%%%%%%%%%%%%%%%%%%%%%%%%%%%%%%%%%%%%%%%%%%%%%%%%%%%%%%%%%%%%%
\section{Results} \label{sec:results}
%It's definitely an active binary which are my favourite class of sources and are much more interesting than boring black holes
\subsection{ATCA J002405.702-720452.361 -- a possible radio counterpart to the X-ray source W286}
\begin{figure*}[ht!]
    \epsscale{0.8}
    \plotone{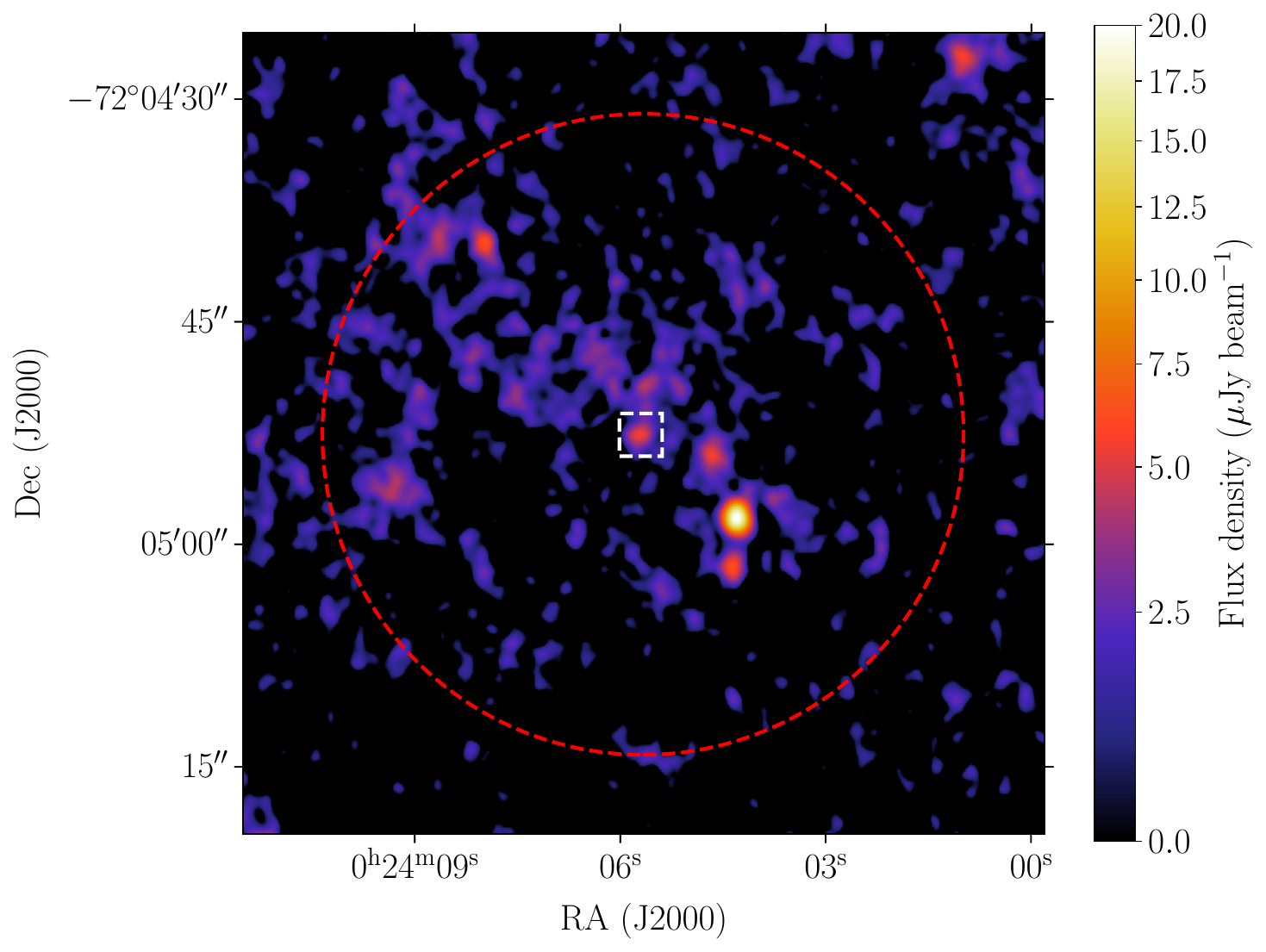}
    \epsscale{0.6}
    \plotone{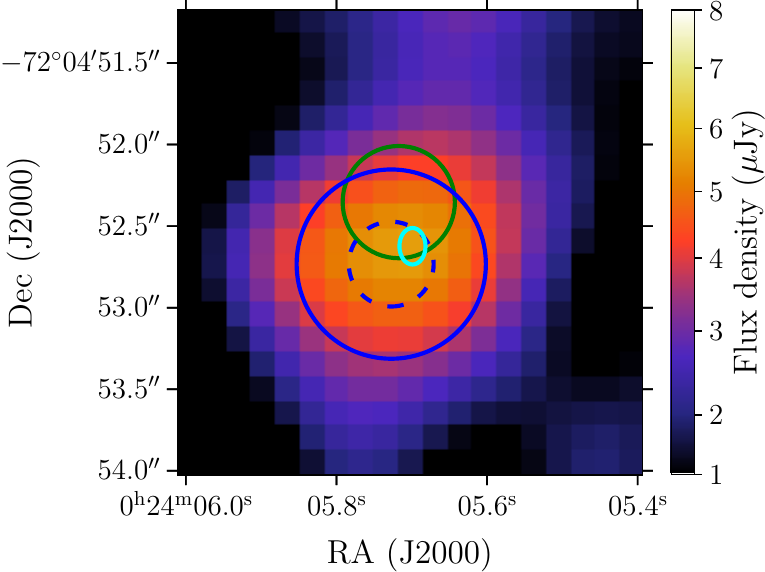}
    \caption{Top: The 7.25 GHz radio image of 47 Tuc. The RMS noise level in this image is $\sim\SI{790}{\nano\jansky\per\beam}$. The core radius of 47 Tuc is shown by the dashed red circle. The white dashed box indicates the field of view of the bottom panel. Bottom: The $1\sigma$ radio position of ATCA J002405 shown in cyan, compared to the $90\%$ X-ray position of the source W286 (green), the cluster centre uncertainty plus the Brownian motion radius added in quadrature for a 6000 M$_{\odot}$ BH (dashed blue circle), compared with the same radius for an illustrative 570 M$_{\odot}$ BH (solid blue circle). All coordinates are in the epoch of the radio image ($\sim$J2021.80). ATCA J002405 is within the uncertainty region of the X-ray source and the cluster centre plus Brownian motion, even at the top end of the possible BH mass range.  \label{fig:core_image}}
\end{figure*}

Our deep ATCA imaging of 47 Tuc revealed a radio source, ATCA J002405.702-720452.361 (ATCA J002405 hereafter), at the photometric centre of the cluster as taken from \citet{Goldsbury2010}. The 7.25 GHz image of the core of the cluster is seen in the top panel of Figure~\ref{fig:core_image}. ATCA J002405 has a 5.5 GHz radio flux density of $6.3\pm1.2$ \si{\micro Jy} and a 9 GHz radio flux density of $5.4\pm0.9$ \si{\micro Jy}. The radio spectral index ($S_{\nu}\propto\nu^{\alpha}$) is $\alpha=-0.31\pm0.54$, and is consistent with being flat. Table~\ref{tab:radio_source} shows the flux density measurements of ATCA J002405 for each of the three main subsets of our campaign, in addition to the full campaign. As the source was not detected in the January subset of the survey, we list the $3\sigma$ flux density upper limits of the source in Table~\ref{tab:radio_source}. The source flux densities were measured using the \textsc{casa} task \verb#imfit# by assuming a point source model, and the $3\sigma$ flux density upper limits were calculated by taking three times the central RMS noise of each image.

The best position (in the epoch J2021.2) of ATCA J002405 is: 
\begin{center}
    $\textrm{RA} = 00:24:05.7018\pm0.0173\ \textrm{s},\quad \textrm{Dec.} = -72:04:52.631\pm 0.112\si{\arcsecond}$.
    \end{center}
This position was derived by using \verb#imfit# to fit a point source model to the radio source, with uncertainties due to the thermal noise of the image. As shown in the bottom panel of Figure~\ref{fig:core_image}, ATCA J002405 is within the positional uncertainty region of the X-ray source W286. Additionally, the angular distance between ATCA J002405 and the cluster centre is \SI{0.14}{\arcsec}, which is less than the sum in quadrature of the Brownian motion radius\footnote{Estimated as $\langle x^2 \rangle=(2/9)(M_{*}/M_{BH}) r_c^2$ where $M_{BH}$ is black hole mass, $M_{*}$ is the average mass of a star in the cluster core (taken to be $\sim1$~M$_{\odot}$) and $r_c$ is the core radius of the cluster. \citep[e.g., see][]{Chandrasekhar1943, Chatterjee2002, Lingam2018}.} for a $6000$~M$_{\odot}$ BH (see Section~\ref{sec:dis_imbh} for details) and the positional uncertainty of the cluster centre, which totals \SI{0.26}{\arcsec}.

ATCA J002405 may also display variability on a timescale of several months. The source is detected in a stacked image in at least one of the frequency bands during the April and September subsets of our survey. It was not detected at $>3\sigma$ in the January subset. This however could be due to slightly higher noise compared to previous subsets (see Table~\ref{tab:radio_source}). Thus we cannot conclusively comment on non-detection being due to intrinsic variability or due to noise fluctuations.

\begin{table}
    \centering
    \caption{The flux density measurements of ATCA J002405 at 5.5, 7.25, and 9 GHz over the course of the survey. The RMS noise for each band is also listed. The chronological subsets were defined based on periods of intense observations in extended configurations, while compact configuration observations were used in imaging in all subsets. The flux density uncertainties also include the uncertainty due to the uncertainty in the calibration. The source was not detected in the January subset of our survey, so we list the $3\sigma$ flux density upper limits in this case.}
    \label{tab:radio_source}
    \begin{tabular}{l|ccc}
        \hline
        \hline
        Subset & Frequency (GHz) & $S_{\nu}$ (\si{\micro\jansky}) & RMS (\si{\micro\jansky\per beam}) \\
        \hline
        Full survey & 5.50 & $6.3\pm1.2$ & 1.19\\
                    & 7.25 & $5.8\pm0.8$ & 0.79\\
                    & 9.00 & $5.4\pm0.9$ & 0.94 \\
        \hline
        April       & 5.50 & $7.8\pm1.9$ &  1.88\\
        (2021-03-31 to 2021-04-23)            & 7.25 & $6.5\pm1.3$ &  1.30\\
                    & 9.00 & $3.7\pm1.5$ &  1.48\\
        \hline
        September   & 5.50 & $5.9\pm1.7$ &  1.71\\
        (2021-06-29 to 2021-10-01)            & 7.25 & $6.5\pm1.1$ &  1.11\\
                    & 9.00 & $7.5\pm1.3$ &  1.35\\
        \hline
        January     & 5.50 & $<6.87$ &  2.29\\
        (2021-11-17 to 2022-01-30)            & 7.25 & $<4.50$ &  1.50\\
                    & 9.00 & $<5.52$ &  1.84\\
        \hline
    \end{tabular}
\end{table}
    
\subsection{X-ray properties of W286}
W286 was first identified as an X-ray source in 2005 \citep{Heinke2005}. To investigate the X-ray spectral properties of W286, we extracted a 0.3-10 keV X-ray spectrum from all the \chandra observations of 47 Tuc and combined them. A circular region with a radius of \SI{1}{\arcsecond} was chosen to extract source counts. An annulus with an inner radius of \SI{1.7}{\arcsecond} and an outer radius of \SI{10}{\arcsecond} was chosen to be the background region, although parts of this annulus were then excluded due to containing other X-ray sources. Source and background spectra were extracted using the \textsc{ciao} task \verb#spec_extract#, and the task \verb#combine_spectra# was used to combine the individual spectra extracted for each observation into one stacked spectrum. Spectral analysis was performed using \textsc{xspec} 12.11.0m \citep{Arnaud1996} and \textsc{bxa} \citep{Buchner2014}. The data were binned to have at least one count per bin, and fitting in \textsc{xspec} used C-stat statistics.

We fit three different models to the data: an absorbed power-law model (\verb#tbabs#$\times$\verb#pegpwrlw#), an absorbed blackbody radiation model (\verb#tbabs#$\times$\verb#bbodyrad#), and an absorbed apec (Astrophysical Plasma Emission Code) model from diffuse ionised gas around the source (\verb#tbabs#$\times$\verb#apec#). For all models, we froze the absorption parameter to the value of the hydrogen column density along the line-of-sight towards 47 Tuc, which is \SI{3.5e20}{\per\centi\metre\squared}. This value is based on an $E(B-V)=0.04$ from the Harris catalogue \citep{Harris1996}, assuming $R_V=3.1$ and the $N_H-A_V$ correlation from \citet{Bahramian2015} and \citet{Foight2016}. From the three models, the power-law model was the best-fitting model, with a photon index of $\Gamma=2.1\pm0.3$, and giving a 0.5-10 keV X-ray luminosity of $2.3^{+0.6}_{-0.5}$\SI{e30}{\erg\per\second}. A blackbody radiation model is the next best-fitting model with a relative probability to the power-law model of 0.79, followed by the apec model with a relative probability of 0.32. The fit parameters are shown in Table~\ref{tab:xray_spectra}. The relative probabilities are calculated in \textsc{bxa}, which uses nested sampling to estimate the model evidence for each model fit, and then calculates relative probabilities.

\begin{table}
    \centering
    \caption{The best-fit parameters of the different X-ray spectral models fit to the combined X-ray spectrum of W286. The photon index ($\Gamma$) is listed in the case of the power-law model, and the electron temperature ($\log(kT)$) is shown in the blackbody radiation and apec model cases. The relative probability indicates the probability of another model being preferred relative to the best-fitting model, which is the power-law model in this case. The relative probabilities are calculated using nested sampling to estimate the model evidence for each model fit.}
    \label{tab:xray_spectra}
    \begin{tabular}{cccc}
        \hline
        \hline
        Spectral fit & $\Gamma$ & $\log(kT)$ (keV) & Relative probability \\
        \hline
        Power-law & $2.1\pm0.3$ & - & 1.00 \\
        Bbodyrad  & - & $-0.5\pm0.1$ & 0.79 \\
        apec      & - & $0.6\pm0.3$ & 0.32 \\
        \hline
    \end{tabular}
\end{table}

\begin{figure*}[ht!]
    \centering
    \includegraphics[width=0.7\textwidth]{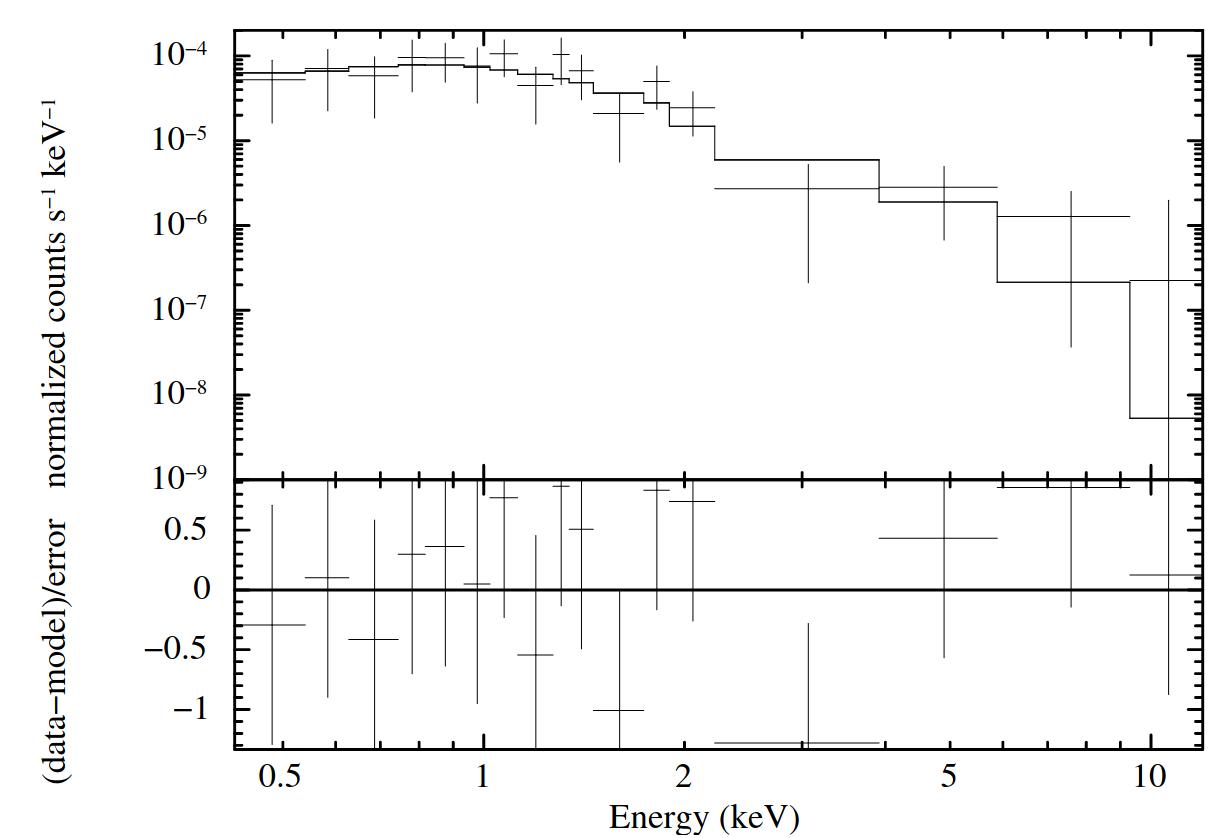}
    \caption{The X-ray spectrum of W286, fit with the best-fitting power-law model in the top panel. The bottom panel indicates the residuals between the model fit and the data. There is no evidence around 1 keV of Fe L-shell emission. \label{fig:xray_spectrum}}
\end{figure*}

We also visually inspected the X-ray spectrum to search for any distinctive features. In particular, we were searching for evidence of Fe L-shell emission, which can be a useful feature in identifying the source class of the object (e.g., ABs). The X-ray spectrum of W286 is shown in Figure~\ref{fig:xray_spectrum}, and also shows the best-fit power-law model and the residuals between the model and the data. As can be seen, there is no evidence of an excess around 1 keV, where the Fe L-shell emission is expected. However, given the low number of source counts, any Fe L-shell emission may be too faint to be detected.

\subsection{Optical and UV properties of W286} \label{sec:results_optical}
\begin{figure*}[ht!]
    \plotone{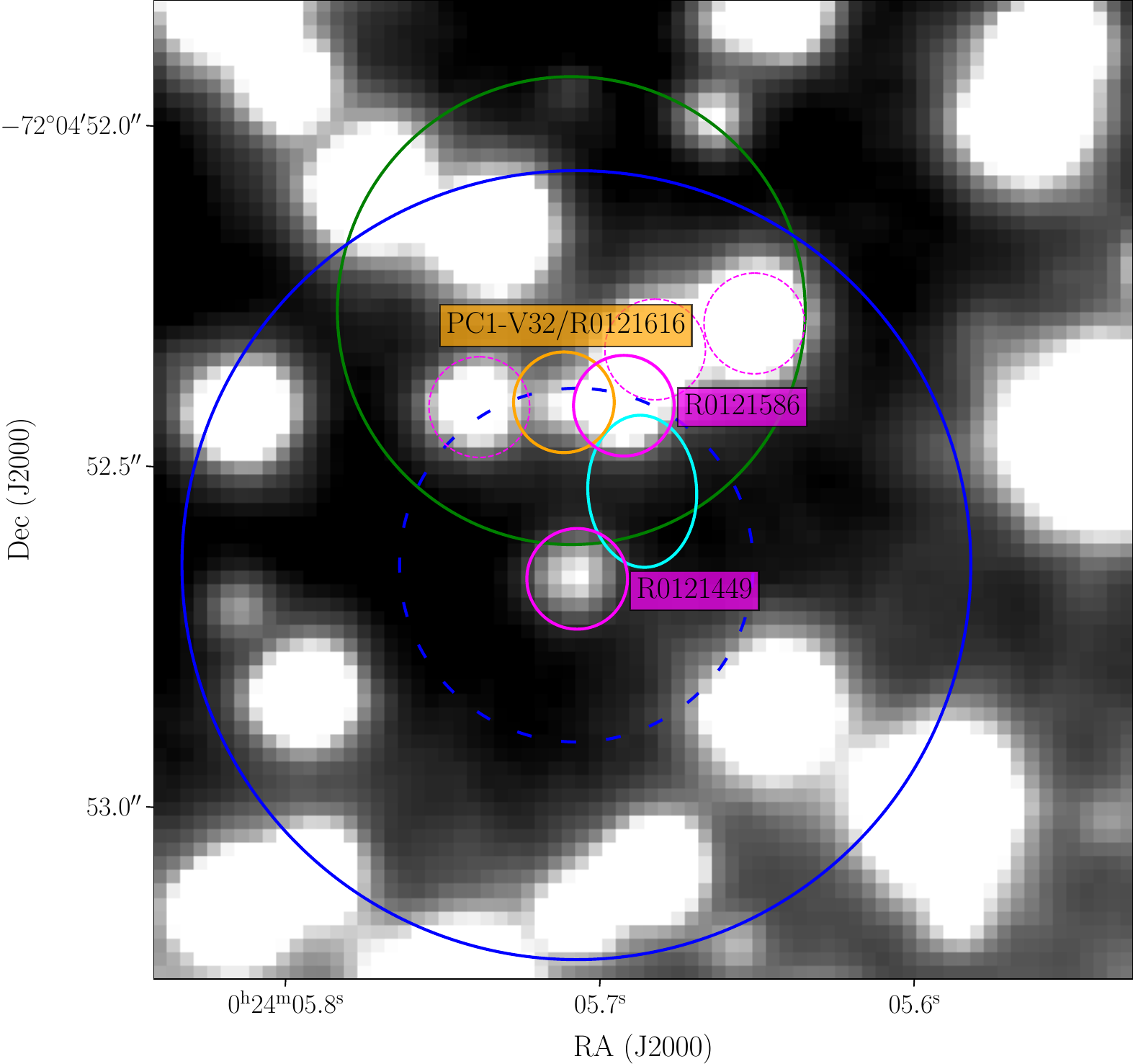}
    \caption{The UV (F300X) HST image of the cluster centre, with positions of the radio source ATCA J002405 (cyan), X-ray source W286 (green), and the cluster centre uncertainty and Brownian motion radius added in quadrature for a 6000 M$_{\odot}$ BH (dashed blue), and the same radius for an illustrative 570 M$_{\odot}$ BH  (solid blue). All circles and ellipses are plotted with coordinates shifted to the epoch of the HST image, J2000. The magenta and orange circles represent positions of optical sources close to ATCA J002405. The solid magenta circles and labels indicate the HUGS sources whose positional uncertainties (represented by the size of the circles) overlap the 1-sigma uncertainty ellipse of the radio source (excluding R0121617 which is unlikely to be real, see \S\ref{sec:results_optical}), with the dashed magenta circles indicating other nearby optical sources. The orange region and label represents the position of PC1-V32, which falls outside the radio uncertainty region. \label{fig:lrs_image}}
\end{figure*}

To identify possible optical counterparts to ATCA J002405, we use a combination of the HUGS survey and the HST image from the F300X filter. Optical sources in the F300X image have a positional uncertainty of \SI{0.074}{\arcsecond} as outlined in \citet{RiveraSandoval2015}. The F300X HST image of the cluster centre is shown in Figure~\ref{fig:lrs_image}. We also consider the BY Dra PC1-V32 as a potential optical counterpart to ATCA J002405 given its previously-claimed association with the X-ray source W286.

From the HUGS survey, we identify three potential HUGS sources whose localization and $1\sigma$ error regions overlap with the $1\sigma$ positional uncertainty of ATCA J002405: R0121449, R0121586, and R0121617. Expanding this to $2\sigma$ includes PC1-V32 and the nearby HUGS source R0121517. PC1-V32 appears to be associated with the HUGS source R0121616 based on the available photometric information of both sources, and is located outside the $1\sigma$ uncertainty region of ATCA J002405. It is important to note that the sources R0121586 and R0121617 appear blended as one source. Upon further inspection, it is likely that the HUGS source R0121617 is not a real source. This source has no cluster membership information in the HUGS catalogue, and was identified in the F435W images in iteration six of the source finding. Upon further inspection of the individual images of the F435W and F336W filters, and the subtracted images in the UV filters, there is no evidence of this source. This, in addition to only one source being detected in the region of the sources R0121617 and R0121586 in the F300X filter, means we are confident that R0121617 is not a real source.

\begin{figure*}[ht!]
\begin{center}
    \includegraphics[scale=0.3,angle=-90]{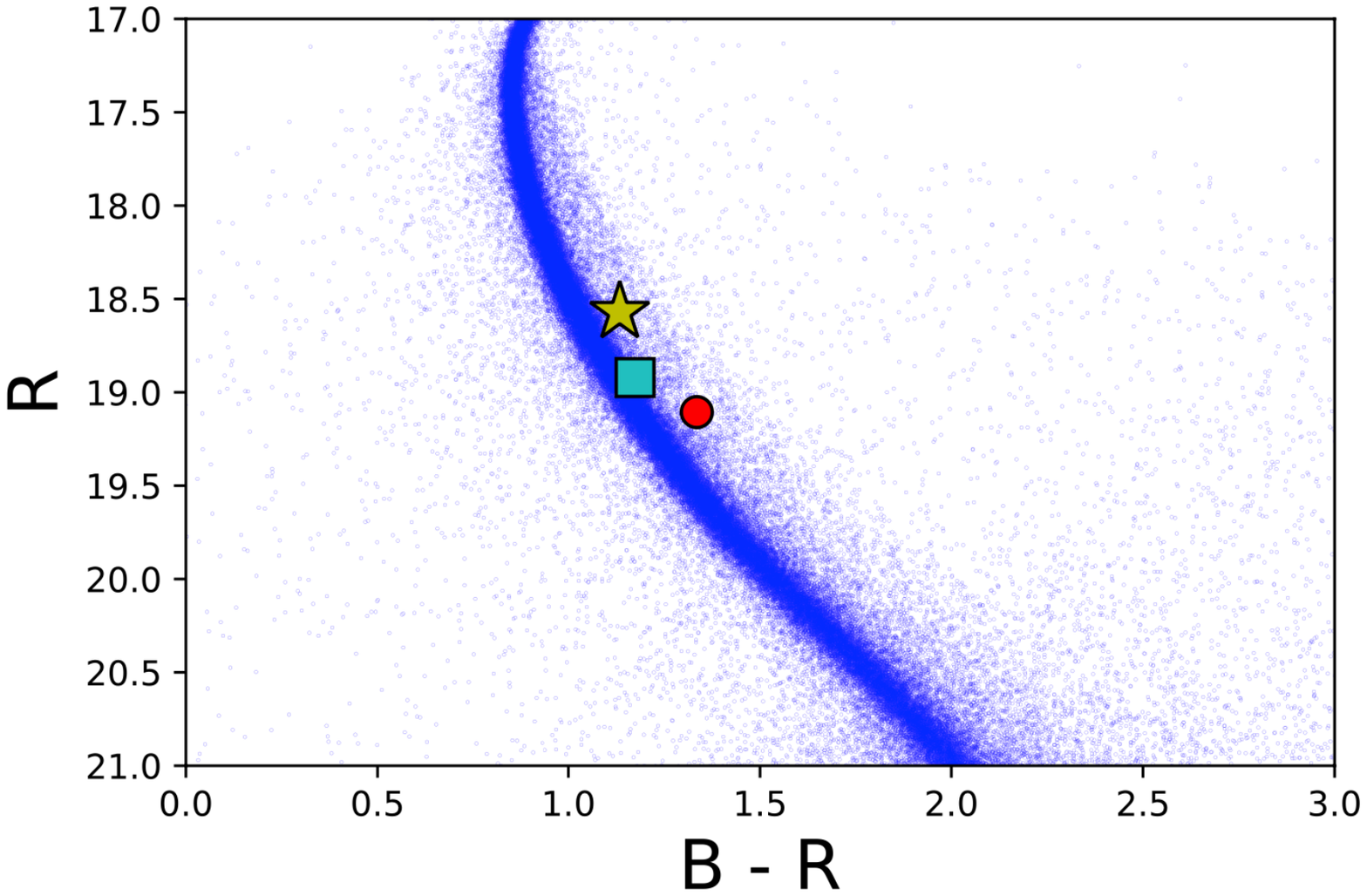}\\
    \includegraphics[scale=0.3,angle=-90]{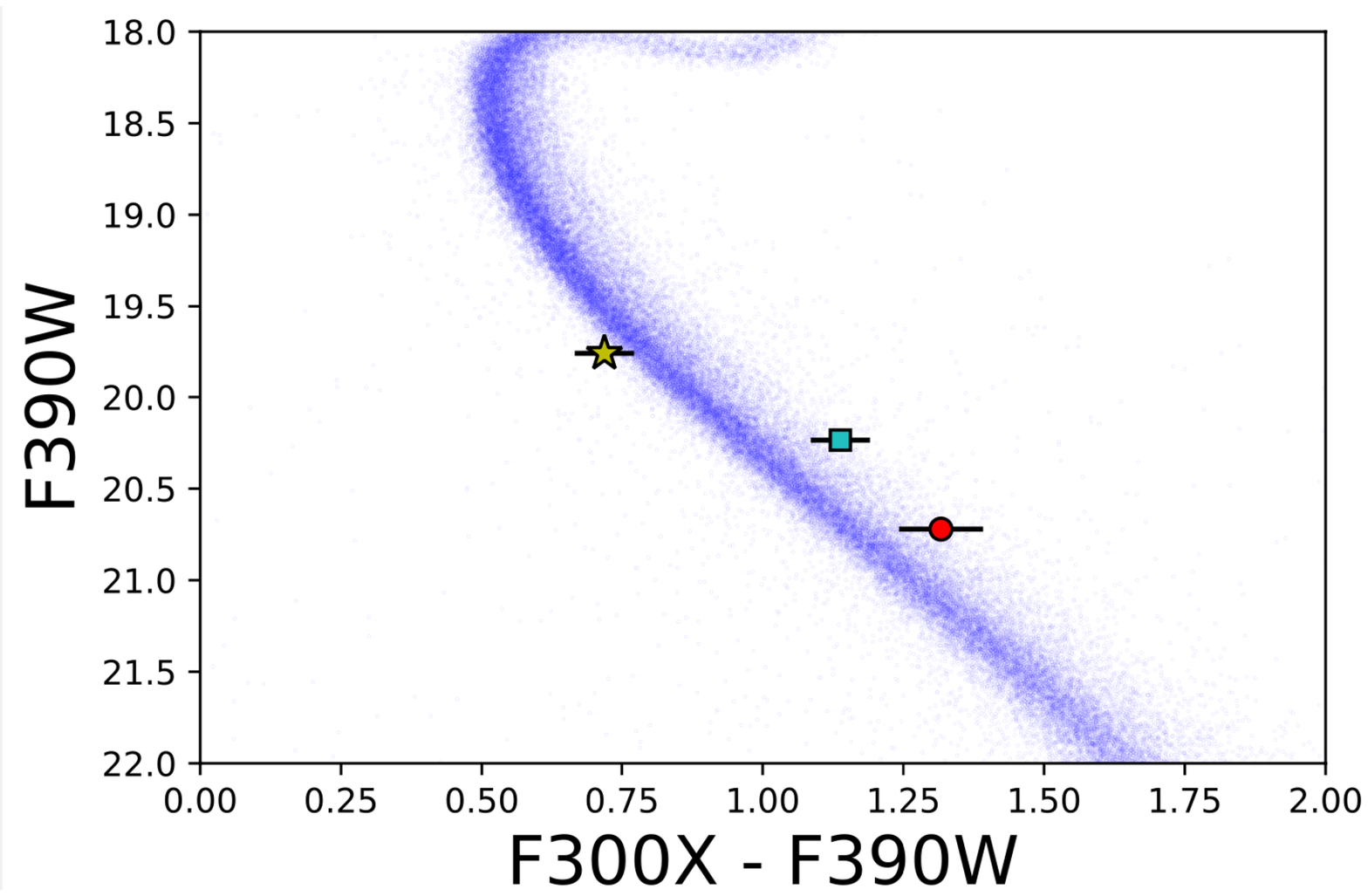}\\
    \includegraphics[scale=0.3,angle=-90]{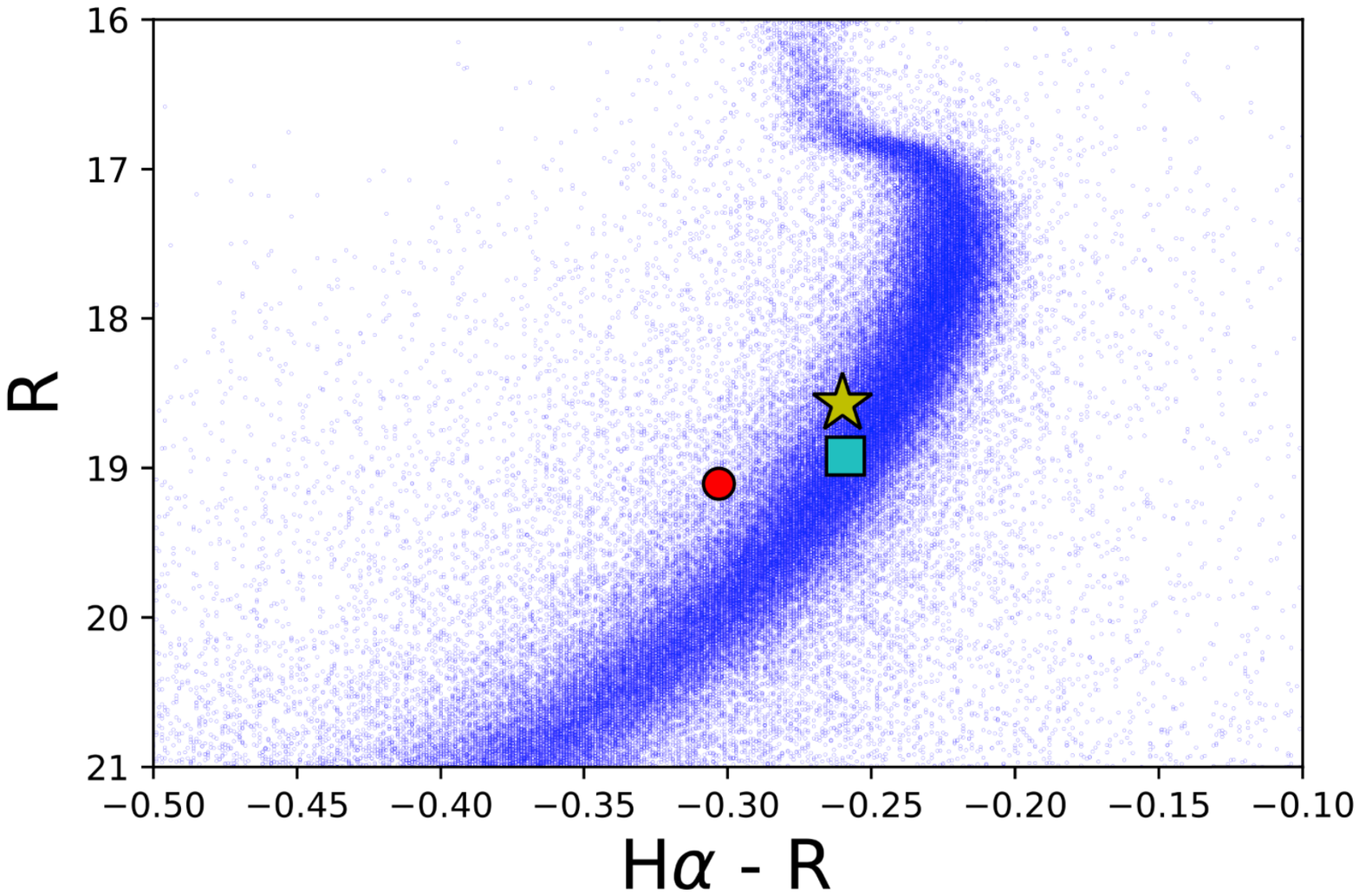}
\end{center}
    \caption{The optical, UV, and H$\alpha$ CMDs of 47 Tuc, with our sources of interest also plotted. The BY Dra PC1-V32 which may be associated with the HUGS source R0121616 is denoted by the red circle, the HUGS source R0121449 is denoted by the cyan square, and the HUGS source R0121586 is denoted by the yellow star. \label{fig:cmds}}
\end{figure*}

To investigate the properties of these optical sources, we constructed colour-magnitude diagrams (CMDs) of 47 Tuc for various combinations of filters. The filter combinations we considered were F390W vs F300X-F390W (UV), $R$ vs H$\alpha$-$R$ (H$\alpha$), and $R$ vs $B$-$R$ (optical). These CMDs are shown in Figure~\ref{fig:cmds}. PC1-V32 (R0121616) shows some H$\alpha$ excess, and appears on the AB sequence in the H$\alpha$ CMD and remains on the binary sequence in the UV CMD. R0121586 appears normal in the optical, UV and H$\alpha$, and R0121449 is is consistent with the main sequence in the optical and H$\alpha$ CMDs, but near the binary sequence in the UV CMD.

\begin{figure*}[ht!]
    \plotone{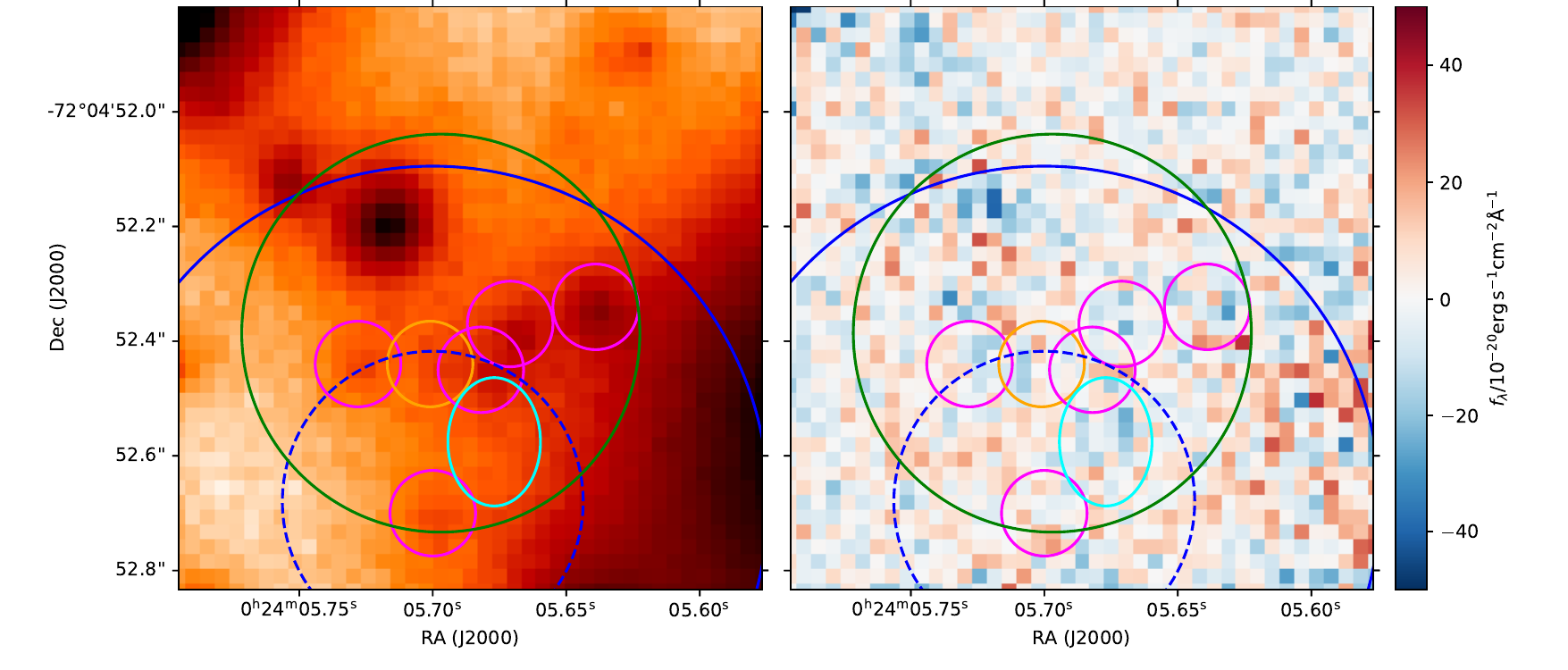}
    \caption{The left panel shows the narrow-band H$\alpha$ image of 47 Tuc using a logarithmic flux map (flux mapped to increase from light colors to dark colors). The circles and ellipses shown in this image are the same as Figure~\ref{fig:lrs_image}. The right panel shows the residuals in H$\alpha$ after the subtraction of stellar spectra, with the linear flux scale indicated by the colorbar. Ellipses and circles are the same as shown in Figure~\ref{fig:lrs_image}. There is no evidence of a resolved H$\alpha$ source in the regions considered. \label{fig:ngc104_muse_halpha}}
\end{figure*}

We also used data from MUSE in narrow-field mode observations to investigate the optical sources in the X-ray uncertainty region. The narrow-band H$\alpha$ image is shown in Figure~\ref{fig:ngc104_muse_halpha}, with the uncertainty regions of the cluster centre, ATCA J002405, W286, and the optical sources indicated. This figure also shows the residual image after the removal of starlight. From this figure, there are no signs of any extended H$\alpha$ emission above the detection limit, and no significant H$\alpha$ source is detected at the location of ATCA J002405.

To complete our optical and UV analysis, we checked to see whether there was a prominent FUV source at the cluster centre that was not present in the optical images. To do this, we visually inspected STIS images of the cluster. This indicated that there was no evidence of any source within our regions of interest down to a magnitude limit of $\sim24$ (the limiting magnitude given in the survey by \citealt{Knigge2002}).

%%%%%%%%%%%%%%%%%%%%%%%%%%%%%%%%%%%%%%%%%%%%%%%%%%%%%%%%%%%%%%%%%%%%%%%%%%%%%%%%%%%%%%%%%%
\section{Discussion} \label{sec:discussion}
\subsection{Possible optical counterparts} \label{sec:disc_1}
We investigated the possible optical counterparts to ATCA J002405, which include the two HUGS sources and PC1-V32 as mentioned in Section~\ref{sec:results_optical}. Based on the available photometric data, PC1-V32 appears to be associated with the HUGS source R0121616. This source appears on the binary sequence in the H$\alpha$ and UV CMDs, which is consistent with it being a BY Dra. PC1-V32 falls outside the $1\sigma$ radio position, as seen in Figure~\ref{fig:lrs_image}, meaning that while it remains possible it is unlikely that the radio emission observed from ATCA J002405 is associated with PC1-V32. We discuss the radio properties of PC1-V32 and other ABs further in Section~\ref{sec:dis_ab}.

We now consider the two other counterparts identified in the HUGS survey, R0121449 and R0121586, as potential counterparts as the uncertainty regions on these sources overlap with the radio uncertainty region of ATCA J002405. R0121449 has typical colours for its brightness in the optical and H$\alpha$ CMDs, and indicates no evidence of accretion. However, the source appears to be consistent within errors with the binary sequence in the UV CMD potentially indicating that it is a binary or variable star. R0121586 is normal in the optical and the H$\alpha$ CMDs, and falls in the scatter of the main sequence in the UV CMD, again showing no strong evidence for accretion. This latter source may also be consistent within errors with the binary sequence, for which there are two possible explanations. R0121586 could be a binary that has a slightly bluish component, or the photometry is affected by crowding and this slight shift to the binary sequence is caused by contamination from the nearby PC1-V32 rather than being intrinsic behaviour of R0121586. These properties indicate that neither of these stars are obvious candidates for the source of the radio emission, and there is no obvious optical counterpart to our radio source. 

\subsection{The origin of the X-ray emission}
We used the MUSE data to investigate the other optical sources (specifically R0121449 and R0121586) within the X-ray uncertainty region to see if there were any glaring features from these sources that could explain the X-ray emission. The H$\alpha$ and residual images are shown in Figure~\ref{fig:ngc104_muse_halpha}. Some of the optical sources have no spectra extracted because they are too faint, including the source that corresponds to PC1-V32. From this figure, there are no signs of any H$\alpha$ emission above the detection limit. This indicates that none of these sources show clear evidence of any accretion present in the system. The extracted spectra of the stars within this region also show no strong evidence for H$\alpha$ emission lines, again indicating that none of these sources are accreting. Overall, none of the optical sources in this field show glaring features in the MUSE data that could favour emission of X-rays via accretion or a similar process. This points to either ATCA J002405 or PC1-V32 being the source of the X-rays.

%%%%%%%%%%%%%%%%%%%%%%%%%%%%%%%%%%%%%%%%%%%
\subsection{Possible radio source interpretations}
\begin{figure*}[ht!]
    \epsscale{0.9}
    \plotone{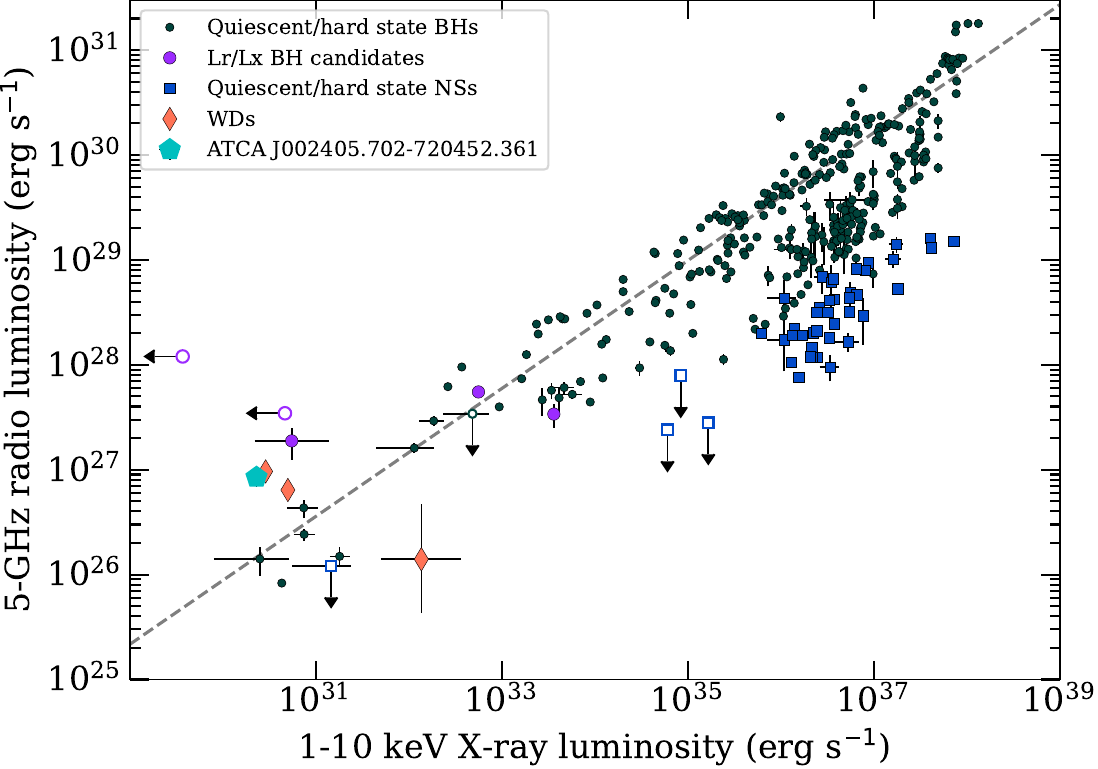}
    \caption{The radio/X-ray luminosity plane for several classes of stellar-mass accreting sources, compiled from \citet{arash_bahramian_2022_6972578}. The y-axis represents the quantity $\nu L_{\nu}$ where $\nu=\SI{5}{GHz}$. The dashed line in this figure indicates the $L_R\propto L_X^{0.6}$ correlation for BHs from \citet{Gallo2014}. Hollow markers with arrows represent upper limits. ATCA J002405 falls well above the standard track for accreting BHs, assuming that it is associated with the X-ray source W286, and assuming a distance of \SI{4.52}{kpc} to 47 Tuc \citep{Baumgardt2021}. The error bars for this point are smaller than the marker size. \label{fig:lrlx}}
\end{figure*}

Given the location of ATCA J002405 in a GC, there are several possible classifications for the source, depending on whether the X-ray emission is associated with the radio source or some other source. By assuming that the X-ray source W286 is associated with ATCA J002405, and a cluster distance of \SI{4.52}{kpc} \citep{Baumgardt2021} we can plot the source on the radio/X-ray luminosity plane of accreting sources (Figure~\ref{fig:lrlx}). ATCA J002405 falls well above the standard track for accreting stellar-mass BHs, and in a part of the parameter space that is occupied by the black hole X-ray binary (BHXB) candidates in M22 and M62 \citep{Strader2012a,Chomiuk2013}, and unusual radio-bright white dwarf (WD) systems. We can use this to identify plausible source classes that could be responsible for this radio emission.

\subsubsection{Direct current (DC) offset in the correlator}
For our observations, we chose our pointing centre prior to 2021 October 1 to coincide with that adopted by all other recent ATCA observations of the data, beginning with \citet{Lu2011}, for ease of stacking the data. However, this raises the possibility of a DC offset in the correlator creating a spurious source at the phase centre.  If two waveforms are multiplied by a DC offset then the offsets can produce a signal, even if there is no correlation between the two signals. 

While a DC offset artefact has not yet been observed in data using the CABB correlator on the ATCA, as soon as we found evidence for a source at the cluster centre we took steps to minimise the possibility that this source could be caused by a system error. For all observations after 2021 October 1 we adjusted the pointing centre of the 47 Tuc scans to the north by \SI{3}{\arcsecond}. An analysis of the stacked data after 2021 October 1 indicated that no source was detected at the pointing centre of the imaged field. This, in addition to private communications with the ATCA Senior Systems Scientist regarding previous experience with the CABB correlator, means we are confident in ruling out a DC offset as the origin of ATCA J002405.

\subsubsection{Active binary} \label{sec:dis_ab}
\begin{figure*}[ht!]
    \plotone{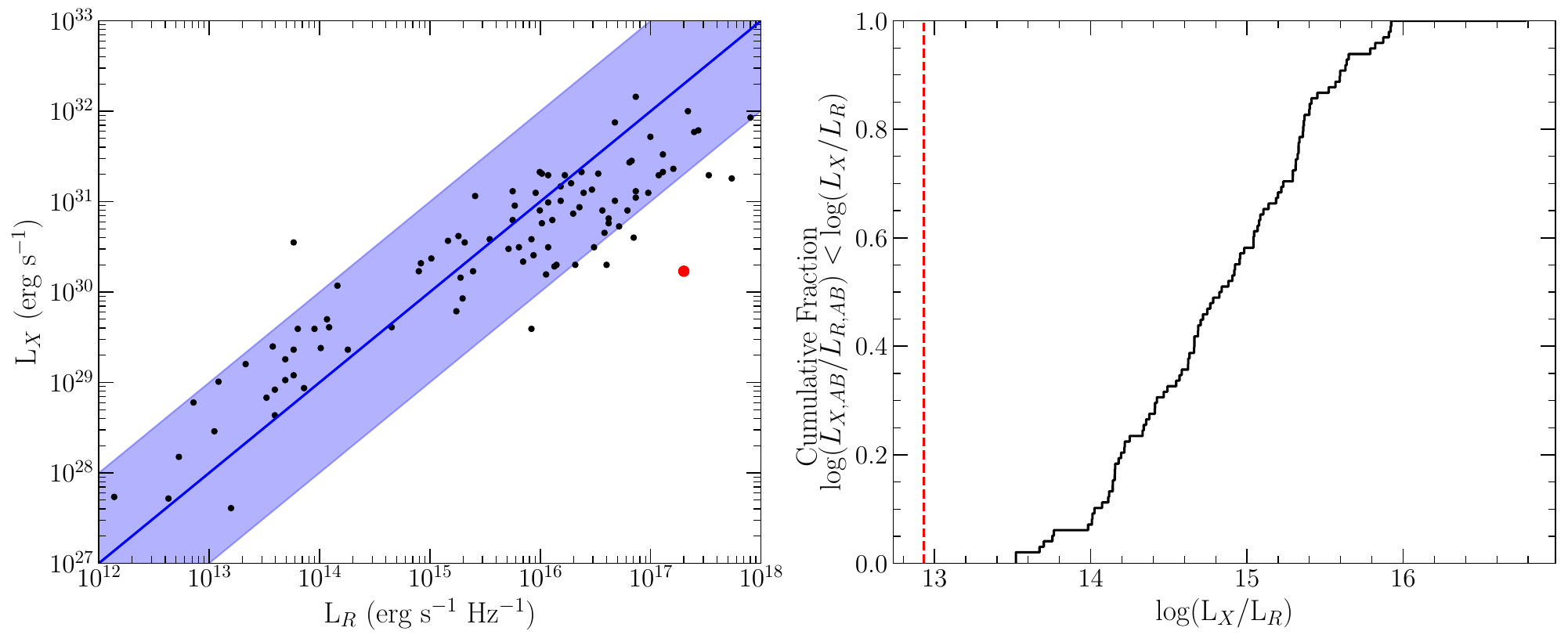}
    \caption{The left panel shows the radio/X-ray relation for active binaries from \citet{Guedel1993}. The red circle indicates our radio detection of ATCA J002405, and that it is outside the scatter of the correlation. The right panel indicates the cumulative fraction of ABs for various $L_X/L_R$ fractions, with the red dashed lined indicating the $L_X/L_R$ fraction of ATCA J002405. \label{fig:AB_LrLx}}
\end{figure*}

ABs, tidally locked stars producing X-ray emission, make up a large portion of the X-ray sources in GCs below $L_X<10^{31}$ \si{\erg\per\second} \citep{Gudel2002}. Furthermore, there is an observed relationship between the radio and X-ray emission of active stars of $L_X/L_R\approx10^{15}$ \citep{Guedel1993}. We find it unlikely that ATCA J002405 is a type of AB. If we assume that the X-ray emission is associated with the radio source, the source becomes a radio-bright outlier on the G\"{u}del-Benz relation, as shown in Figure~\ref{fig:AB_LrLx}, by about an order of magnitude. If the X-ray emission is not associated with the radio source, then the upper limit on the X-ray luminosity will decrease making the source even more of an outlier on the G\"{u}del-Benz relation.

\begin{figure*}[ht!]
    \epsscale{0.5}
    \plotone{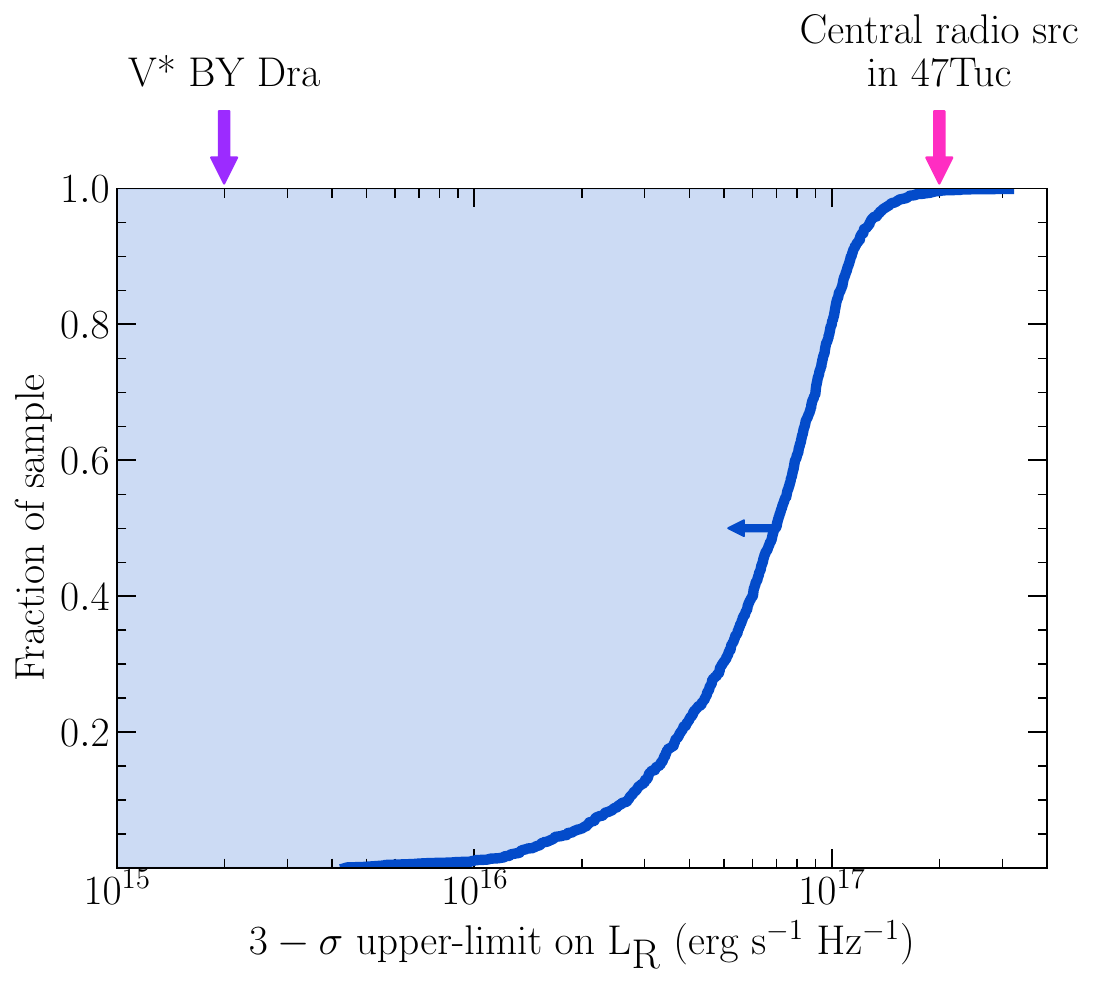}
    \caption{Cumulative distribution of radio luminosity upper-limits for BY Dra variables in the sample investigated in this work. The purple arrow indicates the detected radio luminosity of V* BY Draconis, and the pink arrow indicates the radio luminosity of ATCA J002405 in 47 Tuc. No known BY Dra variable has radio luminosity comparable to ATCA J002405.}
    \label{fig:bydra}
\end{figure*}

Additionally, we have discussed the possibility of the BY Dra PC1-V32 being associated with ATCA J002405. For a comparison to the radio properties of ATCA J002405, we also assessed the radio luminosities of a larger group of BY Dra systems, by constructing a sample based on radio and optical surveys. Our sample of BY Dra was chosen from the Zwicky Transient Facility (ZTF) catalogue of variables \citep{Bellm2019}, as classified by \citet{Chen2020}, who identified a total of 84,697 BY Dra systems. To confine our comparison to the most confident subset of these, we only select sources with at least 200 observations and false alarm probability less than $10^{-5}$ in at least one band. This resulted in a total of 17,015 sources with a confident classification, and periods spanning from 0.15 to 44 days, fully encompassing the 1.64-day period of PC1-V32. We then cross-matched this subset with Gaia DR3 \citep{GaiaCollaboration2022} using a coordinate offset threshold of 0.1$''$, based on the estimated astrometric RMS of the ZTF survey\footnote{Given that Gaia and ZTF surveys have been performed close in time, effects of displacement by high proper-motion are expected to be negligible.} \citep{Masci2019}, yielding 10,237 systems. Of these systems, we retained only sources with parallax significance $>10\sigma$ and located $<0.5$ kpc from Earth, such that wide-area radio surveys such as the Very Large Array Sky Survey \citep[VLASS; typical sensitivity $\sim$128 to \SI{145}{\micro\jansky\per\beam};][]{Gordon2021} would be able to probe radio luminosities to sufficient depth. This resulted in a total of 1,403 BY Dra systems with high confidence classification, tightly constrained distances within 500~pc, and relatively deep constraints on radio luminosity. We then obtained $1'\times1'$ cutout images around each source (using the NRAO VLASS Quick Look database\footnote{\url{https://archive-new.nrao.edu/vlass/quicklook/}}) and searched each of these cutouts statistically (searching for any pixels with peak flux densities above the $3\sigma$ of the $1'\times1'$ cutout) and inspected each visually to verify presence/absence of a source. We found no significant radio sources within 1$''$ of any of these systems. Our $3\sigma$ radio luminosity upper limits are computed from the RMS values of the $1'\times1'$ cutout images for each individual source, and are shown in Figure~\ref{fig:bydra}. We also compare our results to V* BY Draconis (the prototypical BY Dra variable), which is located at 16.5 pc with a consequently high proper motion of $374.74(\pm0.04)$ mas~yr$^{-1}$ \citep{GaiaCollaboration2022}. V* BY Draconis is clearly detected in the VLASS survey at $\sim$\SI{6}{\milli\jansky}. However, this corresponds to a radio luminosity of $\sim2\times10^{15}$ erg~s$^{-1}$~Hz$^{-1}$, over two orders of magnitude radio-fainter than ATCA J002405, further reducing the likelihood that PC1-V32 is the origin of the radio emission. In summary, we find that BY Dra variables are extremely unlikely to show strong radio emission.

\subsubsection{Accreting white dwarf} \label{sec:dis_wd}
Accreting WDs also account for a large number of the X-ray emitting sources in GCs \citep{Grindlay2001,Pooley2002}. Some WDs have been observed to reach the level of radio luminosity that we observe from ATCA J002405. However, these systems are only this radio bright during short radio flares and not persistently. While ATCA J002405 in 47 Tuc may display variability and is undetected in the January campaign of the survey, it was detected in April and September, and CV flares are not as long-lasting as these individual month-long subsets of the survey, or the time between the April and September campaigns. Furthermore, the two optical sources consistent with the radio position of the source show no evidence of being a CV candidate or some other type of radio bright WD as their colours are inconsistent with those of known WDs and CVs in 47 Tuc \citep{RiveraSandoval2018}. Additionally no CV has been identified at this position despite extensive monitoring. Thus, we find it unlikely that ATCA J002405 is a type of CV or other accreting WD. We do note, however, that photometry will be incomplete when going to fainter magnitudes at the centre of 47 Tuc. This means that a faint optical counterpart could still be present, even though it is not seen in the HST data.

\subsubsection{Active galactic nucleus} \label{sec:dis_agn}
\begin{figure*}[ht!]
    \plotone{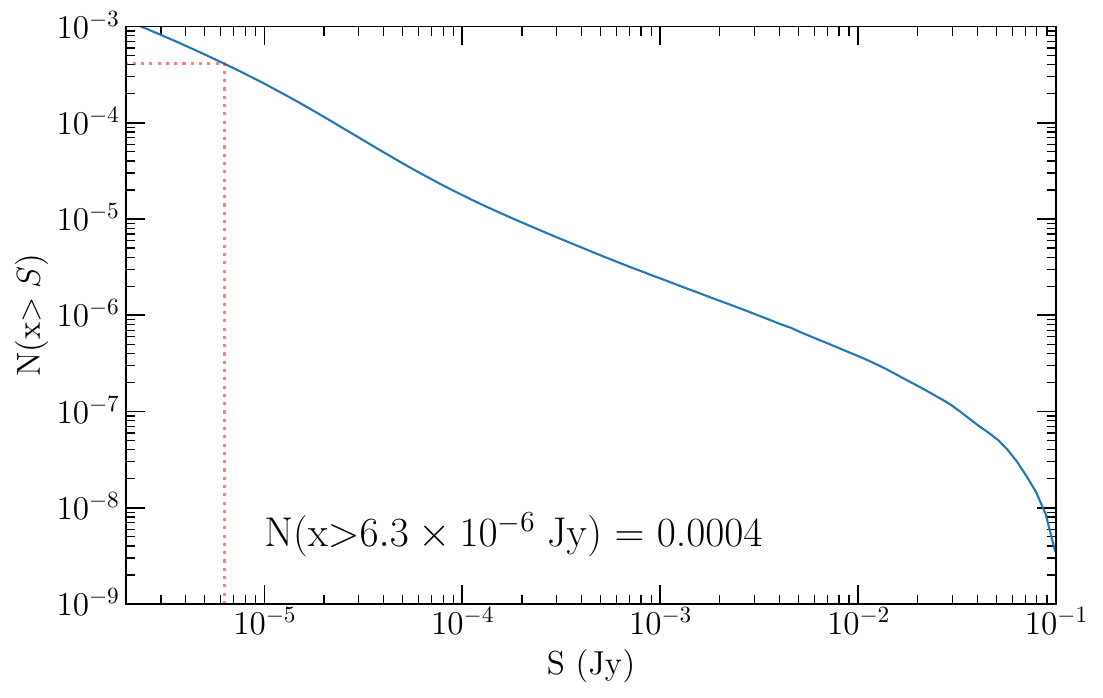}
    \caption{The number of AGN we expect within the sky area of interest with a radio flux density greater than a given threshold plotted against flux density. The red dashed line indicates the 5.5 GHz radio flux density of ATCA J002405. We expect the number of AGN within the sky area of interest with a flux density greater than this to be $\num{4e-4}$, making it unlikely that ATCA J002405 is an AGN. \label{fig:agn_prob}}
\end{figure*}

Active galactic nuclei (AGN) account for a large portion of background sources in the radio sky, meaning that it is entirely possible that ATCA J002405 is an AGN that happens to be coincident with the cluster centre. However, it is unlikely that we would get a background AGN that is coincident with the centre of 47 Tuc. When considering the background differential source counts from \citet{Wilman2008}, the number of AGN expected within one Brownian motion radius of the cluster centre for a $570$~M$_{\odot}$ BH when taking uncertainties into account (\SI{0.47}{\arcsec}) with a radio flux density greater than or equal to that of ATCA J002405 is $\num{4e-4}$. This is shown in Figure~\ref{fig:agn_prob}. Even for a 54 M$_\odot$ BH, the expected number of AGN remains less than $4\times10^{-3}$. This indicates that it is very unlikely that ATCA J002405 is a background AGN. 

\subsubsection{Pulsar} \label{sec:dis_pulsar}
47 Tuc is known for containing a rich collection of pulsars, including millisecond pulsars (MSPs) and other isolated and spider pulsars. To date, there are 29 known pulsars in 47 Tuc\footnote{\url{https://www3.mpifr-bonn.mpg.de/staff/pfreire/GCpsr.html}}. As we will show in a future paper, we have detected several known pulsars in continuum imaging meaning that it is possible we have detected a new pulsar very close to the cluster centre. Such a source may not have been detected in previous pulsar surveys due to its faint radio flux density, and the potential for it to be highly accelerated or hidden around part of its orbit. Spider pulsars, pulsars that are ablating their stellar companion, in particular can be hidden around parts of their orbit due to eclipsing or absorption \citep{Roberts2013}.

\begin{figure*}[ht!]
    \epsscale{0.6}
    \plotone{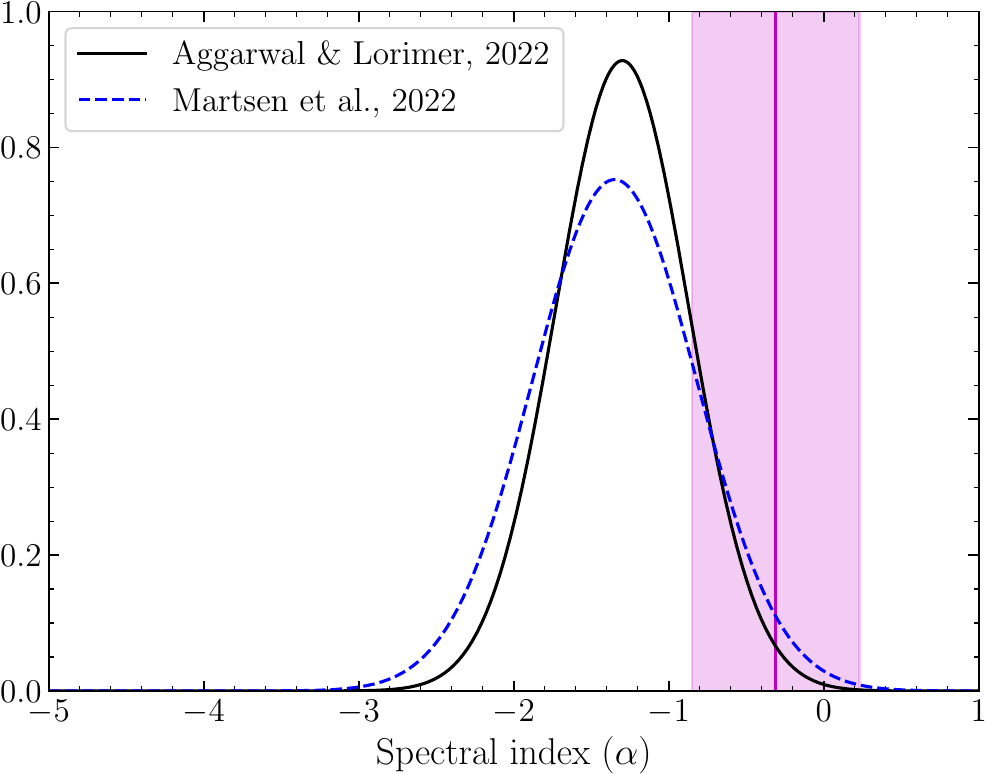}
    \caption{The spectral index distribution for a Galactic sample of MSPs from \citet{Aggarwal2022} shown in black, and that for a sample of MSPs in Terzan 5 from \citet{Martsen2022a} shown in blue. The radio spectral index of ATCA J002405 is shown as the magenta vertical line, with the shaded region indicating the $1\sigma$ uncertainty on this value. The spectral index of ATCA J002405 is consistent with the tail-end of the pulsar spectral index distribution. \label{fig:psr_specindex}}
\end{figure*}

The radio spectral index of ATCA J002405 is $\alpha=-0.31\pm0.54$, which while consistent with a flat spectrum object, is also consistent with the tail-end of the pulsar spectral index distribution. A recent examination of the radio spectra of Galactic MSPs by \citet{Aggarwal2022} have shown that the population of MSPs has a mean spectral index of $-1.3$ with a standard deviation of $0.43$. Further to this, \citet{Martsen2022a} have shown that the MSPs in Terzan 5 have a spectral index distribution with a mean of $-1.35$ and a standard deviation of $0.53$, indicating that the MSPs in GCs seem broadly consistent with the overall MSP population as shown in Figure~\ref{fig:psr_specindex}. The spectral index of ATCA J002405 is consistent with both of these distributions. The probability of obtaining a spectral index $>-0.85$ (the lower uncertainty bound on our spectral index measurement) from the distribution of \citet{Aggarwal2022} is 0.14. Similarly, the probability of obtaining a spectral index $>-0.85$ from the distribution of \citet{Martsen2022a} is 0.17. This indicates that based on its spectral index ATCA J002405 could be an undiscovered pulsar at the centre of the cluster. Further deep observations would be needed by other facilities such as MeerKAT to detect pulsations from and derive a timing solution for this potential pulsar.  For a typical pulsar spectral index of --1.35, its predicted 1.4 GHz flux density would be 40 $\mu$Jy, which while faint is within the range of pulsars previously detected in 47 Tuc \citep{Camilo2000}, while a flatter spectral index, as measured (--0.31), would imply a 1.4 GHz flux density of only $\sim 10 \mu$Jy.

\subsubsection{Stellar-mass black hole} \label{sec:dis_bh}
After decades of debate, we now know that GCs do contain stellar-mass BHs \citep[e.g., ][]{Giesers2018,Giesers2019}. 47 Tuc contains the ultra-compact candidate BH X-ray binary 47 Tuc X9 \citep{Miller-Jones2015,Bahramian2017}, indicating that 47 Tuc may indeed contain some number of stellar-mass BHs. Modelling by \citet{Ye2021} has indicated that 47 Tuc could presently contain around 200 stellar-mass BHs, with a total BH mass in the cluster of $\sim2300$~M$_{\odot}$. Due to the number of BHs expected in 47 Tuc, it is possible that ATCA J002405 is a stellar-mass BH in an XRB that has undergone some brightening. A \chandra DDT observation taken on 2022 January 26 and 27 indicates no increase in X-ray emission from W286 over what was detected in 2015 (albeit this non-detection also corresponds to the radio non-detection in the January subset of the observing campaign), and an analysis of \textit{Swift} X-ray data over the past year shows no increase in total cluster X-ray luminosity, meaning a flaring X-ray transient $>10^{33}$ \si{\erg\per\second} can be ruled out. 

On the radio/X-ray luminosity plane (Figure~\ref{fig:lrlx}), ATCA J002405 falls above the standard correlation for accreting stellar-mass BHs. This is in the same part of the parameter space as the stellar-mass BH candidates in the clusters M22 \citep{Strader2012a} and M62 \citep{Chomiuk2013}, indicating that the radio/X-ray luminosity ratio of ATCA J002405 is similar to other BH candidates in clusters. Other stellar-mass BHs in quiescence have also been shown to occupy this part of the parameter space, specifically MAXI J1348-630 \citep{Carotenuto2022}. The radio spectrum of ATCA J002405 is consistent with being flat, which is what is expected from stellar-mass BHXBs. Furthermore, the best-fitting model to the X-ray spectrum of W286 is a power-law model with a photon index of $\Gamma=2.1\pm0.3$. This means that if W286 is associated with the radio source, its X-ray spectrum is consistent with an accreting BH in quiescence \citep{Plotkin2017}.

While the radio and X-ray properties of the source may be consistent with a stellar-mass BH, we have found no clear optical counterpart to ATCA J002405, although we again note that a counterpart too faint to be detected in HST data could have been missed. As discussed in Section~\ref{sec:disc_1}, the two optical sources consistent with ATCA J002405 show no significant evidence of accretion, or show any significant discrepancy from the single star main sequence within uncertainties.  This is at odds with what has been observed in other stellar-mass BHXB systems. For quiescent BHXBs, we would expect to observe some emission from the outer accretion disk. For example, the disk emission component in the X-ray transient A0620-00 is estimated as at least 10\% of the total light in the near-infrared \citep{Cantrell2010}. Thus, for an accreting BHXB we would not expect the optical counterpart to fall on the main sequence of the cluster CMD. Due to this inconsistency in the optical properties of the source, we cannot conclusively confirm a stellar-mass BH as the classification of ATCA J002405, despite the supportive radio and X-ray behaviours.

\subsection{A central intermediate-mass black hole} \label{sec:dis_imbh}
The multi-wavelength properties of ATCA J002405 do provide evidence that the source could be a candidate IMBH. The position of the source within the radius of Brownian motion of the photometric cluster center (Fig.~\ref{fig:lrs_image}) is where IMBHs in GCs would be expected to be located - they are the heaviest objects in the cluster and should have migrated to the cluster centre by dynamical friction. The radio spectrum of the ATCA J002405 is also flat, consistent with what is expected for an IMBH in a low-luminosity state. Although ATCA J002405 does show evidence of variability, some radio variability is expected in quiescence \citep{Plotkin2019} and has been observed previously in other BH systems, such as V404 Cygni and Sgr A*. 

We can make a rough estimate of the mass of any potential IMBH through the fundamental plane of BH activity. For consistency with previous works \citep[e.g.][]{Tremou2018}, we adopt the following form of the fundamental plane with mass as the dependent variable as shown in \citet{Miller-Jones2012}:
\begin{equation} \label{eq:fp}
\begin{split}
    \log M_{BH} &= (1.638\pm0.070)\log L_R - (1.136\pm0.077)\log L_X\\ &- (6.863\pm0.790),
\end{split}
\end{equation}
where the BH mass is in solar masses, and the radio and X-ray luminosities are in \si{\erg\per\second}. We note that the uncertainties in the fit parameters are strongly correlated, such that standard error propagation of the terms in Equation~(\ref{eq:fp}) would overpredict the uncertainty on black hole mass estimates. 

For the case where the X-ray emission from W286 is associated with the radio emission from ATCA J002405, we can get a direct estimate of the BH mass from the radio and X-ray luminosities. The 5.5 GHz flux density of the source is \SI{6.3}{\micro\jansky}, corresponding to a 5.5 GHz luminosity of \SI{8.47e26}{\erg\per\second} at a distance of 4.52 kpc \citep{Baumgardt2021}, and the 0.5-10 keV X-ray luminosity of W286 is then \SI{2.3e30}{\erg\per\second}. This corresponds to a position on the fundamental plane of $570^{+430}_{-260}$~M$_{\odot}$, where the quoted uncertainty is purely statistical, and was calculated by considering only the errors on the radio and X-ray luminosities. However, the scatter on the fundamental plane will add further systematic uncertainty to this mass estimate. The most robust examination of the intrinsic scatter in the fundamental plane was performed by \citet{Gultekin2019}. Through Markov Chain Monte Carlo simulations, \citet{Gultekin2019} found a systematic uncertainty in this relation of $\SI{0.96}{dex}$. It is this more conservative estimate of $\sim$\SI{1}{dex} that we take to represent the intrinsic scatter in the fundamental plane. This means that our mass estimate of any BH is uncertain by at least a further order of magnitude, giving a nominal $1\sigma$ mass range of $\sim54-6000$ M$_{\odot}$.

Due to this scatter in the fundamental plane, our mass estimate cannot currently be made any more certain. It appears that the scatter in the mass direction for this relation is largely driven by supermassive black holes, and it is unknown how this scatter translates to lower-mass BHs such as IMBHs, which represent unexplored parts of the fundamental plane parameter space. Overall, despite its large intrinsic scatter, the fundamental plane still remains the only currently available method to estimate the mass of this source.

Within the cluster center, the Brownian motion radius for a black hole can be estimated as $\langle x^2 \rangle=(2/9)(M_{*}/M_{BH}) r_c^2$, where $M_{*}$ is the average mass of a star in the cluster core (taken to be $\sim1$~M$_{\odot}$) and $r_c$ is the core radius of the cluster\footnote{ $r_c$ assumed to be 0.61~pc for 47 Tuc, based on \citet{Baumgardt2021}.} (see \citealt{Strader2012} and \citealt{Tremou2018}). For the black hole mass constraint of 54--6000~$M_{\odot}$, the resulting Brownian motion radii would be in the range 0.17--2.0$''$, with the smallest Brownian motion radius corresponding to the highest mass. The source position lies within the combined uncertainty (summing in quadrature the Brownian motion radius and the uncertainty on the cluster centre) of the cluster centre, even for the smallest Brownian motion radius of 0.17$''$.

The mass estimate above only considers the case where both the radio emission from ATCA J002405 and the X-ray emission from W286 are associated with a central IMBH. It is important to also consider the case where the X-ray emission from W286 is not associated with ATCA J002405 and is instead associated with PC1-V32, in which case we can again use the fundamental plane to estimate a BH mass lower-limit. This can be calculated using the measured radio luminosity of the source, and an upper-limit on the X-ray luminosity from the source. Given that the radio source position falls within the uncertainty region of W286, we adopt the X-ray luminosity of W286 as the X-ray luminosity upper-limit. This limit would lie at the same point on the fundamental plane relation as the direct measurement above, so in this case we adopt the BH mass lower limit (accounting for the scatter in the fundamental plane) of $54$~M$_{\odot}$. It is worth noting that a BH with a mass of $54$ M$_{\odot}$ will have a Brownian radius of $\approx2''$. This is large enough that the probability density of the BH being located so close to the cluster center would be low. This tentatively argues against the mass being quite this low.

It is also important to consider the case where both the radio emission from ATCA J002405 and the X-ray emission from W286 are not associated with a candidate IMBH. Because the of the presence of a radio source within the Brownian motion region, it is possible that the radio emission from a potential IMBH could be hiding in the wings of the radio emission from ATCA J002405. In this case, we can use the flux density of ATCA J002405, an estimation of the X-ray emission expected from an IMBH accreting from the intra-cluster gas, and the fundamental plane relation to derive a mass upper-limit for a central IMBH in 47 Tuc. This process follows the methodology outlined in \citet{Strader2012} and \citet{Tremou2018}. The X-ray luminosity of a source is related to its accretion rate ($L_{X}=\epsilon\dot{M}c^2$), and for accretion rates less than 2\% of the Eddington rate the radiative efficiency $\epsilon$ scales with accretion rate \citep{Maccarone2003,VahdatMotlagh2019} and can be expressed as:
\begin{equation}
    \epsilon = 0.1\qty( \frac{\dot{M}}{\dot{M_{Edd}}}/0.02).
\end{equation}

The accretion rate $\dot{M}$ is assumed to be some fraction of the Bondi accretion rate, usually $\sim0.03$ \citep{Pellegrini2005,Maccarone2007}. To compute the Bondi accretion rate we take the gas number density to be $n=0.2$ cm$^{-3}$, consistent with pulsar measurements \citep{Abbate2018}. We also assume that the gas is fully ionised with a temperature $T=\SI{e4}{\kelvin}$ and a mean molecular mass $\mu=0.59$ which is typical for fully ionised gas \citep{Fall1985}. For consistency with \citet{Strader2012} and \citet{Tremou2018}, we only consider the $\gamma=1$ isothermal case, although we note that the $\gamma=5/3$ adiabatic case will result in a higher mass upper-limit.

Using the flux density of ATCA J002405 (\SI{6.3}{\micro\jansky}), the upper limit would then sit at $860$~M$_{\odot}$ on the fundamental plane relation. This is lower than the corresponding fundamental plane estimate calculated by \citet{Tremou2018}, due to the unparalleled image depth that we have achieved. However, to translate this to a true BH mass limit we must again account for the intrinsic scatter in the fundamental plane relation, giving an upper limit on the BH mass of $7900$ M$_{\odot}$.

%%%%%%%%%%%%%%%%%%%%%%%%%%%%%%%%%%%%%%%%%%%%%%%%%%%%%%%%%%%%%%%%%%%%%%%%%%%%%%%%%%%%%%%%%%
\section{Conclusion} \label{sec:conclusion}
In this paper, we present the deepest radio image of the globular cluster 47 Tucanae. Our ultra-deep imaging campaign with the Australia Telescope Compact Array at 5.5 and 9 GHz has allowed us to reach an RMS noise level of \SI{790}{\nano\jansky\per\beam}, representing the deepest radio continuum image made of a globular cluster and the deepest radio image ever made by the Australia Telescope Compact Array.

Based on analysis of these data, we have identified ATCA J002405.702-720452.361, a flat-spectrum, variable radio source that falls within the uncertainty region of the faint X-ray source W286 and the cluster centre. This source has a 5.5 GHz flux density of $6.3\pm1.2$ \si{\micro\jansky} and a spectral index of $\alpha=-0.31\pm0.54$. We consider several possible explanations for the origin of the radio and X-ray emission, and conclude that the radio source does not originate from the previously proposed counterpart to W286, a BY Draconis source (PC1-V32), and that the X-ray emission from W286 is associated with either PC1-V32 or this newly discovered radio source. We consider several possible source class explanations for ATCA J002405.702-720452.361, which are summarised in Table~\ref{tab:conclusion}, and we find that the most likely classifications for the source are either an undiscovered pulsar or an intermediate-mass black hole. A stellar-mass black hole appears less likely than these other explanations but cannot be ruled out.

\begin{table*}
    \centering
    \caption{Comparison of explanations for the origin of the radio emission.}
    \label{tab:conclusion}
    \begin{tabular}{lcl}
        \hline
        \hline
        Source class & Possibility & Comment \\
        \hline
        Active binary & ? & Radio emission is too bright for an active binary at this distance (Section~\ref{sec:dis_ab})\\
        White dwarf & ? & Colours of optical counterparts are inconsistent with known WDs (Section~\ref{sec:dis_wd})\\
        AGN & ? & Unlikely to have an AGN at this flux density this close to the cluster centre (Section~\ref{sec:dis_agn}) \\
        Stellar-mass BH & \ding{51} & $L_R/L_X$ ratio can be consistent with other BH candidates, no clear optical counterpart (Section~\ref{sec:dis_bh}) \\
        Pulsar & \ding{51}\ding{51} & Spectral index is consistent with pulsar spectral index distribution (Section~\ref{sec:dis_pulsar})\\
        IMBH & \ding{51}\ding{51} & Radio source is located at the cluster centre, mass estimates $54-6000$~M$_{\odot}$ (Section~\ref{sec:dis_imbh})\\
        \hline
    \end{tabular}
\end{table*}

It is not unsurprising that a pulsar at the cluster centre may have been missed. ATCA J002405.702-720452.361 is very faint, meaning that it may have been invisible to previous low-frequency pulsar surveys of 47 Tucanae. Additionally, the pulsar could be highly accelerated or hidden for parts of its orbit, either through absorption or eclipsing by a binary companion, again rendering it invisible. Further deep observations with MeerKAT would be needed to fully explore this possibility, in addition to potential higher or lower-frequency detections to attempt to further constrain the spectral index of the source. These observations would allow us to test whether ATCA J002405.702-720452.361 is another member of 47 Tucanae's large pulsar population.

In the event that ATCA J002405.702-720452.361 is an intermediate-mass black hole, we use the fundamental plane of black hole activity to estimate the mass of the source. In the event that the radio and X-ray emission are associated, the source location on the fundamental plane would sit well below previous upper limits on BH mass. Accounting for the $\sim 1$\,dex intrinsic scatter on the fundamental plane, we find a $1\sigma$ uncertainty range on the BH mass of $54-6000$~M$_{\odot}$. This can be further reduced at the upper end by considering the kinematic studies of the central region of the cluster \citep[][see \S \ref{sec:intro47tuc}]{DellaCroce2023}, to give a mass range of $54-578$~M$_{\odot}$. An intermediate-mass black hole with mass $\leq578$~M$_{\odot}$ would have a sphere of influence of $\leq\SI{0.8}{\arcsecond}$. A better mass estimate for any central BH could be achieved if an orbiting companion is identified, allowing for the mass to be dynamically measured. It would be valuable to get proper motions for as complete as possible a sample of stars within this sphere of influence, either with ground-based adaptive optics data or with the James Webb Space Telescope. This would allow the presence of a central intermediate-mass black hole to be tested, either through dynamical modelling or through searching for a potential companion star.

\begin{acknowledgments}
We would like to thank the anonymous referee for their useful comments on the manuscript. AP would like to thank Jamie Stevens and the other support staff and duty astronomers responsible for the operation of the Australia Telescope Compact Array for their help and assistance during this observing campaign. Additionally, AP would like to thank the director of the \chandra X-ray Observatory, Patrick Slane, for approving a request for Director's Discretionary Time, and the support staff at the \chandra X-ray observatory for their assistance in performing this observation. AP would also like to acknowledge Freya North-Hickey and Ben Quici for useful discussions about data reduction procedures, Chris Riseley and Cyril Tasse for useful discussions involving imaging and self-calibration, Cath Trott for useful discussions regarding error propagation, and Maureen van den Berg for useful discussions regarding photometry. COH thanks Greg Sivakoff for useful discussions. 

AP was supported by an Australian Government Research Training Program (RTP) Stipend and RTP Fee-Offset Scholarship through Federation University Australia. SK acknowledges funding from UKRI in the form of a Future Leaders Fellowship (grant no. MR/T022868/1). JS acknowledges support from NASA grant 80NSSC21K0628 and the Packard Foundation. COH is supported by NSERC Discovery Grant RGPIN-2016-04602. 

The Australia Telescope Compact Array is part of the Australia Telescope National Facility which is funded by the Australian Government for operation as a National Facility managed by CSIRO. We acknowledge the Gomeroi people as the traditional owners of the Observatory site.

\end{acknowledgments}


\vspace{5mm}
\facilities{Australia Telescope Compact Array, \chandra X-ray Observatory, Hubble Space Telescope, MUSE}


\software{
Astropy \citep{AstropyCollaboration2018}, 
CASA \citep{McMullin2007}, 
CIAO \citep{Fruscione2006}, 
DDFacet \citep{Tasse2018},
Miriad \citep{Sault1995}, 
Matplotlib \citep{Hunter2007},
Numpy \citep{Harris2020}
}



\bibliography{references}{}
\bibliographystyle{aasjournal}



\end{document}